\documentclass[twocolumn,twocolappendix]{aastex631}
\usepackage[utf8]{inputenc}
\usepackage{amsmath, mathtools}
\usepackage{graphicx}
\usepackage{wrapfig}
\usepackage{float}
\usepackage{natbib,color,ulem,hyperref}

\makeatletter
\renewcommand\paragraph{\@startsection{paragraph}{4}{\z@}%
            {-2.5ex\@plus -1ex \@minus -.25ex}%
            {1.25ex \@plus .25ex}%
            {\normalfont\normalsize}}
\makeatother
\def\refnew#1{(\ref{#1})}
\def\erg{\,\rm erg}
\def\g{\,\rm g}
\def\K{\,\rm K}
\def\yr{\,\rm yr}
\def\s{\,\rm s}
\def\km{\,\rm km}
\def\kms{\,\rm km\,s$^{-1}$}
\def\cm{\,\rm cm}

\def\bpic{$\beta$\,Pic}

\newcommand{\uu}[1]{#1} 
\newcommand{\yy}[1]{#1}

\newcommand{\w}{\color{blue}}
\newcommand{\y}{\color{magenta}}

\shorttitle{Argon and Debris Disks}
\shortauthors{Wu et al.}

\begin{document}

\title{Argon in $\beta$ Pictoris -- entrapment and release of volatile in disks}

\author[0000-0003-0511-0893]{Yanqin Wu}
\affiliation{Department of Astronomy \& Astrophysics, University of  Toronto, 50 St George Street, Toronto, ON M5S\,3H4, Canada}
\author[0000-0002-5885-5779]{Kadin Worthen}
\affiliation{William H.\ Miller III Dept.\ of Physics and Astronomy, John’s Hopkins University, 3400\,N.\ Charles Street, Baltimore, MD\,21218, USA}
\author[0000-0002-7201-7536]{Alexis Brandeker} 
\affiliation{Department of Astronomy, Stockholm University, AlbaNova University Center, 106\,91 Stockholm, Sweden}
\author[0000-0002-8382-0447]{Christine Chen}
\affiliation{William H.\ Miller III Dept.\ of Physics and Astronomy, John’s Hopkins University, 3400\,N.\ Charles Street, Baltimore, MD\,21218, USA}
\affiliation{Space Telescope Science Institute, 3700 San Martin Drive, Baltimore, MD\,21218, USA}


\begin{abstract}
Chemical compositions of planets reveal much about their formation environments. Such information is well sought-after in studies of Solar System bodies and extra-solar ones. Here, we investigate the composition of planetesimals in the $\beta$~Pictoris debris disk, by way of its secondary gas disk. We are stimulated by the recent JWST detection of an \ion{Ar}{2} emission line, and aim to reproduce extensive measurements from the past four decades.  Our photo-ionization model reveals that the gas has to be  heavily enriched in C, N, O, and Ar (but not S and P), by a uniform factor of about 100 relative to other metals. Such an abundance pattern is both reminiscent of, and different from, that of Jupiter's atmosphere. The fact that Ar, the most volatile and therefore the hardest to capture into solids, is equally enriched as C, N, and O suggests that the planetesimals were formed in a very cold region ($T \leq \uu{20}- 35\K$), \uu{possibly with the help of entrapment if} water ice \uu{is over-abundant}.
In the debris disk phase, these volatile are preferentially out-gassed from the dust grains, likely via photo-desorption. The debris grains must be `dirty'
 aggregates of icy and refractory clusters.
Lastly, the observed strength of the \ion{Ar}{2} line can only be explained if the star \bpic\ (a young A6V star) has sizable chromospheric and coronal emissions, on par with those from the modern Sun.
In summary,  observations of the \bpic\ gas disk rewind the clock to reveal the formation environment of planetesimals. 
%
%
%
\end{abstract}

\section{The Motivations}

The recent JWST detection of the \ion{Ar}{2} 6.986\,$\mu$m line \citep{Worthen2024} reignited our interest in the gas disk around $\beta$~Pictoris. 
It comes as a surprise that Argon, a noble gas with a lower abundance than many other elements, should emit the brightest line in the mid-infrared (MIR). Our investigation into this surprise uncovers interesting information. 

At a distance of  $\sim 19$\,pc, and spatially extended ($\sim 100$\,au), the \bpic\ debris disk is a favorite object for many studies. Both its gas and dust components have been
extensively mapped \yy{\citep[e.g.,][]{Chen2007, Cataldi2018, Matra2017,Rebollido2024,Worthen2024}}, to study the origin of debris disks \uu{\citep[see reviews by][]{Wyatt2008,Hughes2018}}. The gas component, likely sourced from the dust grains, \uu{also} yields kinematic information unavailable if only the dust component is visible. 

The gas in \bpic\ and other debris disks also opens a profitable window into another issue, chemical composition of the parent planetesimals. As such information is intimately related to the formation history, it has been actively sought after. For example, for Solar system bodies, the abundances of volatile gases and noble gases are often measured by expensive space missions \citep[see reviews by][]{Hersant2004,depater2023}.
In debris disks with gas components, these elements are in plain view and can be measured with the simple act of taking a spectrum.



\begin{table}
\hskip-0.35in
\begin{tabular}{l|cl|c}
\hline
atom/ion/ & measured& references & model\\
\uu{molecule} & [$\cm^{-2}$]  &  & [$\cm^{-2}$]\\
 \hline
H$^{0}$ & $< {\rm a\, few}\times 10^{19}$ & [11] & $4.5\times 10^{17}$\\
H$^{+}$ &   &   & $6.5\times 10^{17}$ \\
C$^{0}$ & $3\times 10^{16}$ \textdagger & [2] & $7.2\times 10^{16}$\\ 
C$^{+}$  & $2\times 10^{16}$ \textdagger & [5] & $2.3\times 10^{17}$\\
C$^{++}$ & $1.3\times 10^{13}$ & [3]&  $9.5\times 10^{12}$\\
N$^{0}$  & $8\times 10^{14}$ \textdagger & [4] & $4.2\times 10^{16}$ \\
N$^{0}$ & {$8\times 10^{16}$} & {[this work]} & $4.2\times 10^{16}$ \\
N$^{+}$ &   & & $3.4\times 10^{16}$\\
O$^{0}$  & $5\times 10^{15}$ \textdagger  & [5] & $2.9\times 10^{17}$ \\

O$^{+}$ & &  & $2.6\times 10^{17}$\\
Ar$^{0}$ & & & $1.3\times 10^{15}$ \\
Ar$^{+}$ & & & $1.5\times 10^{15}$\\
Ne$^{0}$ & & & $9.5\times 10^{14}$\\
Ne$^{+}$ & & & $7.0\times 10^{12}$\\
Na$^{0}$ & $4.5\times10^{10}$ & [10] & $4\times 10^{10}$\\
Na$^{+}$ & $> 3\times 10^{13}$ & [5] & $2.0\times 10^{13}$ \\
Mg$^{0}$ & $2.5\times 10^{11}$ & [6] & $1.3\times 10^{12}$ \\
Mg$^{+}$ & $> 2\times 10^{13}$ & [7] & $4.4\times 10^{14}$\\
Al$^{0}$ & $< 4\times 10^{10}$ & [7] & $6.5\times 10^7$\\
Al$^{+}$ & $4.5\times 10^{12}$ & [7] & $3.2\times 10^{13}$\\
Si$^{0}$ & $< 10^{13}$ & [7] & 
$1.4\times 10^{12}$\\
Si$^{+}$ & $10^{14}$ & [7] & 
$3.6\times 10^{14}$\\
P$^{0}$ & $<7\times 10^{11}$ &[5] & $5.8\times 10^{10}$\\
P$^{+}$ & $<9.2\times 10^{13}$&[5] & $2.8\times 10^{12}$\\
S$^{0}$ & $5.4\times 10^{12}$ & [7] & $2.1\times 10^{12}$\\
S$^{+}$ &    & & $1.4\times 10^{14}$ \\
S$^{++}$ & $<4\times 10^{11}$ & [4] & $1.7\times 10^{10}$ \\
Ca$^{0}$ & $3\times 10^8$ & [8] & $4.7\times 10^8$\\
Ca$^{+}$ & $1.26\times 10^{13}$ & [9] &$2.3\times 10^{13}$ \\
Ca$^{++}$ & & & $1.9\times 10^{12}$\\
Cr$^{0}$ & & & $9.8\times 10^{9}$\\
Cr$^{+}$ & $3.5\times 10^{12}$ & [7] & $4.6\times 10^{12}$\\
Mn$^{0}$ & $<3\times 10^{10}$ & [7]  & $5.4\times 10^9$\\
Mn$^{+}$ & $3\times 10^{12}$ & [7] & $3.0\times 10^{12}$\\
Fe$^{0}$ & $10^{12}$ & [7] & $4.0\times 10^{11}$\\
Fe$^{+}$ & $3.7\times 10^{14}$ & [5] & $3.7\times 10^{14}$\\
Ni$^{0}$ & $< 7.6\times 10^{10}$ & [5] & $2.9\times 10^{10}$\\
Ni$^{+}$ & $1.5\times 10^{13}$ & [7] & $1.8\times 10^{13}$\\
Zn$^{0}$ & $< 7\times 10^{10}$ & [7] & $3.7\times 10^{10}$ \\
Zn$^{+}$ & $2\times 10^{11}$ & [7] & $3.7\times 10^{11}$\\ 
\uu{CO}& \uu{$6.3 \times 10^{14}$} & \uu{[2]} & \\
 \hline
\end{tabular}
\caption{Column densities for \uu{various species}, both measured from absorption line studies (error-bars suppressed) and calculated in our model.  \textdagger\ marks those obtained from unresolved UV absorption lines. They may be plagued by saturation and should be considered as lower limits.
References: [1]~\citet{Wilson2017}; [2]~\citet{Roberge2000}; [3]~\citet{Bouret2002}; [4]~\citet{Wilson2019}; [5]~\citet{Roberge2006}; [6]~\citet{vidal1994}; [7]~\citet{Lagrange1998}; [8]~\citet{Kiefer2019}; [9]~\citet{Crawford1994}; [10]~\citet{vidal1986};[11]~\citet{Freudling1995}. We provide an updated N$^{0}$ determination in Appendix~\ref{sec:N_data}.}
\label{tab:elements}
\end{table}

With the initial goal of explaining the Ar line, we carry out a study of the \bpic\ gas disk. Our results serve to answer questions on two main directions: one, origin of gas in the \bpic\ disk; and what do their abundance pattern reveal about the process of planet formation. We also provide new JWST analysis of a few emission lines (Appendix \ref{sec:app1}).

\subsection{Origin of gas in the \bpic\ debris disk}
\label{sec:origin}

Ever since its discovery four decades ago \citep{Smith1984}, the debris disk around \bpic\ has been under intense study. But \uu{much of the disk,} especially its gas component, remains enigmatic to date. 
In the same way as the observed $\mu$m-sized dust must be recently generated, since their collisional lifetime is much less than the system age \citep{Backman1993}, so must the gas be secondary in origin.  Observed to be collocated with the dust \citep{Brandeker2004,Nilsson2012}, they are not remnant from the proto-planetary disk but are rather released from the orbiting debris  \uu{\citep{Wilson2017,Matra2017,Cataldi2018,Cataldi2023}}.
Various hypothesis on how the gas is released have previously been presented: through dust sublimation \uu{\citep[perhaps aided by radioactive decay,][]{Bonsor2023}}, through collisional vaporization \citep{Czechowski2007}, through photo-desorption of icy particles \citep{Artyomowicz1997,Chen2007,Grigorieva2007}, \uu{through escape from inclusions during collisional fragmentation \citep{Zuckerman2012}},
or through a combination of re-accretion and release \citep{Cataldi2020}. 

Not only are we unsure how the gas is released, we are also uncertain about when.
\uu{The system lifetime is pegged at $\sim 20$\,Myr \citep{Mamajek2014}. And the disk as observed in  \ion{Na}{1} \citep{Brandeker2004} is radially extended and smooth, consistent with having undergone viscous diffusion. However,} 
the spatial distribution of CO \uu{\citep{Dent2014,Matra2017}} and C$^0$ \citep{Cataldi2018} show a \uu{pronounced} north-east/south-west asymmetry that implies a recent ($< 5000$\,yr) increase in CO gas production \citep{Cataldi2018}, as the C$^0$ otherwise should have azimuthally sheared into a smooth ring and later diffused into an accretion disk \citep{Kral2016}. Recent observations by JWST/MIRI of dust emission shows a curved extension bending away from the disk and that seems connected to the CO asymmetry, implying a perhaps even more recent gas production event (100--200\,yr) for the asymmetric component \citep{Rebollido2024}.  Both these estimates are much shorter than the system lifetime. 

At least one puzzle on the gas disk is resolved. It has long been wondered why refractory elements such as Na, Ca and Fe, 
which experience strong radiation pressure \citep{Beust1989}, appear to orbit the star in nearly Keplerian orbits \citep{Olofsson2001, Brandeker2004}. \citet{Fernandez2006} proposed that an excess of C, by a factor of $10$, is sufficient to brake the ions by Coulomb interactions. The C over-abundance in the gas disk was subsequently confirmed by \citet{Roberge2006} and \citet{Cataldi2014}, which naturally resolved the issue \citep[see also][]{Zagorovsky2010}.  In this work, we find that C, N, and O are actually enriched by a factor of $100$, rendering radiation pressure completely unimportant. This also answers the question posed by \citet{Xie2013}: is the volatile enrichment caused by `preferential production' of volatiles, or `preferential depletion' of the refractories by radiation pressure? The answer is the former.
%

\subsection{Observational constraints}
\label{sec:observables}

The four decades of \bpic\ observations have yielded a wealth of data, which we will compare our model against.

As the disk of \bpic\ is seen edge on, its gas content was detected in absorption against the star \citep{Slettebak1975} long before the discovery of the disk \citep{Smith1984}. The initial assumption was that the gas is located within 1\,au of the star \citep{Lagrange1998}, but later spatially resolved observations of \ion{Na}{1} showed the emission to be co-located with the dust disk and the gas in orbits consistent with Keplerian velocities \citep{Olofsson2001}.

The edge-on geometry is a great advantage when constraining the column densities of various species, as using the star as a background light source enables very sensitive measurements of resonant transitions. In particular in the UV, many elements have strong transitions from the ground state. Table~\ref{tab:elements} shows column densities of some 30 atomic and ionic species, inferred from absorption line studies.

Column densities do not directly constrain where along the line of sight the absorbers are located. Complementary studies of spatially resolved \uu{emission scattered} from resonance lines in atoms and ions \citep[in particular from \ion{Na}{1}, \ion{Fe}{1}, and \ion{Ca}{2}, see][]{Olofsson2001,Brandeker2004,Nilsson2012} were therefore important to constrain the spatial extent of the \uu{gas} disk.

Since the disk gas is expected to be relatively cold (Fig.~\ref{fig:structure}), only low-energy fine-structure and molecular states are expected to be collisionally excited, with transitions in the infrared to sub-millimeter. See Table~\ref{tab:lines} for detected fine-structure emission lines from the disk gas. Among these, the \ion{C}{1} 610\,$\mu$m line was spatially resolved \citep{Cataldi2018}. Table~\ref{tab:lines} also includes 3 new lines from JWST that are published here for the first time (Appendix \ref{sec:app1}). We also present the spatial information we obtain from JWST on these lines in  Table \ref{tab:resolved_flx}.

Here, we focus on atomic (and ionic) species as molecules are rapidly dissociated. \uu{While some CO is observed in the disk\citep[e.g.,][]{Roberge2000, Dent2014}, it is less important} for the overall thermal/chemical budget. We also only consider the so-called `stable' gas of the main disk -- gas at heliocentric velocities $\sim 20$\kms\ (\bpic's stellar velocity) -- and not the temporally variable gas absorption features that have been interpreted as Falling Evaporating Bodies \citep[FEBs, ][]{Beust1989}.

\defcitealias{Wilson2019}{W19}

Many resonant transitions in the UV  are so strong that they become heavily optically thick. In combination with unresolved complex line profiles, it becomes difficult to constrain the column densities with high accuracy, as the observed, unresolved line becomes a non-linear function of the actual line profile. An example is the determination of the O column density 
\citep{Roberge2006} that was later found to likely have been underestimated \citep{Brandeker2016}. A similar instance concerns the column density of \ion{N}{1} \citep[hereafter \citetalias{Wilson2019}]{Wilson2019}, where the 1200\,\AA\ triplet as observed with HST/COS was found to be strongly optically thick and best fit with multiple unresolved line components. 
Since their derived constraint implied a N$^{0}$ column density that lies about two orders of magnitudes below our prediction (Fig.~\ref{fig:elements}), we decided to take a closer look to better quantify the upper limit on the N column density. Our study is presented in Appendix~\ref{sec:N_data}. We conclude that our predicted overabundance of N is indeed consistent with the HST/COS data, and maybe even preferred as the derived line profile better matches the observed absorption lines.


\begin{table*}
\hskip-1.0in
\begin{tabular}{l|c|c|c|l|c|c}
\hline
cooling line & observed flux & spectral res. & spatial res. & references & model flux &   \uu{$\tau_{\rm line}$}\\
 &  [$\erg/\s/\cm^2$]  & [km/s] & &  & [$\erg/\s/\cm^2$] & \uu{(radial)} \\
 \hline
\ion{C}{2} 158\,$\mu$m & $(2.4\pm 0.1)\times 10^{-14}$ & $0.17$
& no
&\citet{Cataldi2014} & $2.2\times 10^{-14}$  & \uu{$0.65$}
\\
\ion{C}{1} 610\,$\mu$m & $(1.6\pm 0.2)\times 10^{-16}$ & $0.34$
& yes
&\citet{Cataldi2018} & 
\uu{$2.2\times 10^{-16}$}
& \uu{$1.2$}\\
\ion{O}{1} 63.2\,$\mu$m & 
$(1.24 \pm 0.25)\times 10^{-14}$ & $86$  & no & \citet{Brandeker2016} & 
\uu{$5.6\times 10^{-15}$}
& \uu{ $4.3$} \\
\ion{Ar}{2} 6.99\,$\mu$m & $2.4{ \pm0.1}\times 10^{-14}$& $88$ & yes & \citet{Worthen2024} & $2.6\times 10^{-14}$ & \uu{$1.2\times 10^{-2}$}\\
\ion{Ne}{2} 12.81\,$\mu$m & $< 4\times 10^{-15}$ & $88$& - & this work  & 
$3.4\times 10^{-17}$ & \uu{$3.0\times 10^{-5}$}
\\
\ion{Fe}{2} 17.93\,$\mu$m   & $4.3{ \pm0.5} \times 10^{-15}$& $88$ & yes & this work & $1.6\times 10^{-15}$
& \uu{$2.9\times 10^{-4}$}
\\
\ion{Fe}{2} 25.99\,$\mu$m   & $1.1 { \pm0.1}\times 10^{-14}$ & $88$ & yes & this work & $9.4\times 10^{-15}$ & \uu{$1.4\times 10^{-2}$}\\ 
\hline
\end{tabular}
\caption{Atomic cooling lines detected from the \bpic\ gas disk.  See Appendix \ref{sec:app1} for some new measurements. \uu{The last column reports the line radial optical depths  from our model ($\tau_{\rm line}$).}
}
\label{tab:lines}
\end{table*}

\subsection{Volatiles and Planet Formation}
\label{subsec:intro2}

\uu{Volatile elements in the gas disk place constraints on the formation of their parent planetesimals. As will become clear, the \bpic\ disk is highly enriched in volatiles C, N, O and Ar. }
%
%
These elements are also sometimes called super-volatiles, as their main carriers in the proto-planetary disks (CO, N$_2$ and Ar) are notoriously difficult to be captured into solid bodies.
Their existence in planets and moons thus provides a natural thermometer for the formation environment. In the following, we briefly review the processes of volatile capture, before giving a short re-cap of Solar system measurements that are pertinent to our study. 

There are 3 known ways to capture volatiles: direct condensation; 
entrapment by porous water ice \citep{Bar-Nun1988}; and clathrate hydrate formation in crystalline water ice \citep{Lunine1985}. 
%
Direct condensation of Ar, N$_2$ and CO require extremely low temperatures \citep[see, e.g.,][]{Iro2003,Hersant2004}. \uu{The condensation temperature for the most volatile of all, Ar, lies around $20\K$, for nebular pressure and abundances.
If we adopt a stellar luminosity of $L_* = 4 L_\odot$ during the disk phase, and estimate the mid-plane temperature in a passively irradiated proto-planetary disk following \citet{Chiang1997},}
\begin{equation}
T_{\mathrm{mid-plane}} \approx {35\K} \, \left(L_*\over{4\,\mathrm{L_\odot}}\right)^{1/4} \, 
\left({r\over{65\,{\rm au}}}\right)^{-3/7}\, ,
    \label{eq:Tcg97}
\end{equation}
\uu{we find that direct condensation may be difficult within the observed disk.}

In contrast to direct condensation, entrapment by water ice (either by porosity or through clathration) enables volatile capture at higher temperatures, ranging from \citep{Iro2003,Bar-Nun1988,Greenberg2017,Simon2023} 45\K\ for CO (the most stable volatile), to 40\K\ for N$_2$, and to 35\K\ for Ar (the hardest to trap). 
At temperatures below $35\K$, all hyper-volatiles can be indiscriminately trapped.


This entrapment process depends \uu{not only on temperature, but also on} the abundance of water. 
As more stable volatiles (e.g., CO and N$_2$) take up available host sites first \citep{Lunine1985,Iro2003}, a shortage of water may lead to a strong depletion of the more volatile gases (e.g., Ar). 
We can estimate the amount of water required, in terms of the ratio $N_{\mathrm{O}}/N_{\mathrm{C}}$. 
As each CO or N$_2$ molecule requires $5-6$ H$_2$O molecules as hosts 
\citep{Hersant2004}, complete \uu{C and N capture (which then allows Ar to be entrapped)}  suggests that  $N_{\mathrm{O}}/N_{\mathrm{C}} \geq 
\uu{6.6}
$ (for a solar ratio of $N_{\mathrm{C}}/N_{\mathrm{N}} = 4$).\footnote{\uu{This value is lowered slightly to $6.25$ if most of nitrogen is instead in the form of NH$_3$. One water molecule is required per NH$_3$ hydrate \citep{Hersant2004}.}}  This is \uu{almost 4 times higher} than the solar value of  $(N_{\mathrm{O}}/N_{\mathrm{C}})_\odot \sim 1.8$. 



\uu{In contrast, volatile capture by direct condensation preserves the primordial O/C ratio. These considerations make the O/C ratio a useful probe of the formation process.} 

A number of volatiles have been measured in Jupiter's atmosphere, by ground-based observations and space missions like Voyager, Galileo and Juno \citep[for a recent review, see][]{depater2023}.\footnote{There are no reported measurements of the refractories, likely because they have condensed out from Jupiter's cold atmosphere. }
It appears that Jupiter is uniformly enriched in C, N,  S, P, Ar, Kr, and Xe, by about a factor of 3 \citep{Atreya1999}.
This uniform enrichment, including even the most volatile Ar, suggests that the solids that are responsible for enriching Jupiter's atmosphere 
are formed at a very cold locale in the solar system, \uu{at least} $T < 35\K$ \citep{Owen1999,Oberg2019}, \uu{and as cold as $20\K$ if direction condensation dominates.}
This is much cooler than that expected for a passively irradiated disk at 5\,au \citep[$T\sim 75\K$, eq. \ref{eq:Tcg97} but applying to the Sun,][]{Chiang1997} and remains a puzzle \citep{Owen1999,Oberg2019}.
\uu{Measurements of the O abundance have been notoriously difficult, due to water condensation in the planet's atmosphere \citep{Wong2008}. According to a recent theoretical inference by \citet{Li2020}, O is also enriched by $\sim 3$ times, making direct condensation a likely scenario. However, we note that \citet{Li2024} recently updated their inference on O abundance to 4.9 times solar, with a possible range 1.5--8.3. While this is higher than before, it still falls below that expected of volatile entrapment by water ice (only),} where one requires a 
 minimum oxygen over-abundance of
\begin{eqnarray}
&  & {{(N_{\mathrm{O}}/N_{\mathrm{H}})}\over{(N_{\mathrm{O}}/N_{\mathrm{H}})_\odot}}   = 
{{(N_{\mathrm{O}}/N_{\mathrm{C}})(N_{\mathrm{C}}/N_{\mathrm{H}})}\over{
(N_{\mathrm{O}}/N_{\mathrm{H}})_\odot}}\nonumber \\
& & = 
{{ 
\uu{6.6}\times 3\times (N_{\mathrm{C}}/N_{\mathrm{H}})_\odot}\over{(N_{\mathrm{O}}/N_{\mathrm{H}})_\odot}} = {{
\uu{6.6}
\times 3}\over {1.8}}\sim 
\uu{11}
\, .
\label{eq:jupiter}
\end{eqnarray} 
\uu{So for Jupiter, the origin of its volatiles remains unclear, for now.} 


Measurements on other Solar system bodies are even more sparse. Existing measurements on Saturn, Uranus and Neptune suggest that they are also heavily enriched in the volatile elements (C, S, and P). Measuring their abundances in elements like Ne, Ar, Kr and Xe remains to be a key objective of future space missions \citep[e.g.,][]{Atreya2018,Simon2018}.
Objects at larger distances, e.g., Titan (a moon of Saturn) and Oort cloud comets, are surprisingly depleted in Ar and/or N \citep{Niemann2005,Balsiger2015}, in contrast to 
the cold condition inferred for Jupiter, making for a confusing interpretation.

%

Given this state, data from other systems are much welcomed. As we will show in this work,
gas observations in the \bpic\ debris disk open a new window towards understanding volatile reservoirs in outer planetary systems.


\subsection{This Work}

Our main tool here is the CLOUDY spectral synthesis code \citep{Ferland2013}. We will construct a model that  contains two major ingredients: first, the gas disk (gas radial distribution, abundance pattern), and second, the star itself. Although \bpic\ is a bright main-sequence A6V star, its radiation is far from being well understood. Unexpected high UV and X-ray emission of unclear origin has been observed from the star. These high-energy radiations dramatically affects the ionization balance in the gas disk, and hence its emission/absorption properties. We will introduce our model for the star and the gas disk in \S \ref{sec:star}--\ref{sec:diskmodel}, respectively; in \S \ref{sec:comparisons}, we use CLOUDY to infer the elemental abundances that are consistent with observational constraints. 
We then discuss the implications of our results, both for the origin of gas in debris disks (\S \ref{sec:discussions}), and for the elemental abundances of planetesimals (\S \ref{sec:abundances}).

\section{Model for the Stellar Emission}
\label{sec:star}

We adopt a photospheric model of $T_{\rm eff}=8052\K$ interpolated from the ATLAS model \citep{Castelli2003}, with a bolometric luminosity of $8.7\,L_\odot$ \citep{Crifo1997}. Since the metallicity of \bpic\ is found to be insignificantly different from solar \citep[{[Fe/H]} = 0.11 $\pm$ 0.1,][]{Saffe2008}, we assume it to be exactly Sun-like.

In addition to the photospheric radiation, we assume that the star contains two more components that are important for the gas disk: a hot, tenuous corona, and a denser, somewhat cooler chromosphere below the corona. 
\uu{The chromosphere in particular endows the star with a much higher UV flux over its blackbody radiation, motivating our omission of the comparatively small interstellar UV contribution. }

\subsection{Chromosphere model}

Argon has a first ionization potential of 15.7\,eV, at  which the photospheric flux is negligible. 
So a hot chromosphere is essential to explain the observed \ion{Ar}{2} emission (and also the \ion{C}{2} line). In addition, continuum UV photons from the chromosphere are important for heating the gas disk. 

Empirically, gas hotter than the photosphere has also been known to exist on \bpic, as evidenced by the strong emission lines in Ly\,$\alpha$ and \ion{O}{5} \citepalias{Wilson2019}, and \ion{C}{3}, \ion{C}{4}, and \ion{O}{6} \citep{Bouret2002}. These lines have equivalent widths of a few hundred \kms. The luminosity in the Ly\,$\alpha$ line alone
amounts to a few $\times 10^{-5}\,L_\odot$
\citepalias{Wilson2019}. 

\begin{figure}
\centering
\hspace{-0.05in}
\includegraphics[width=0.5
\textwidth]{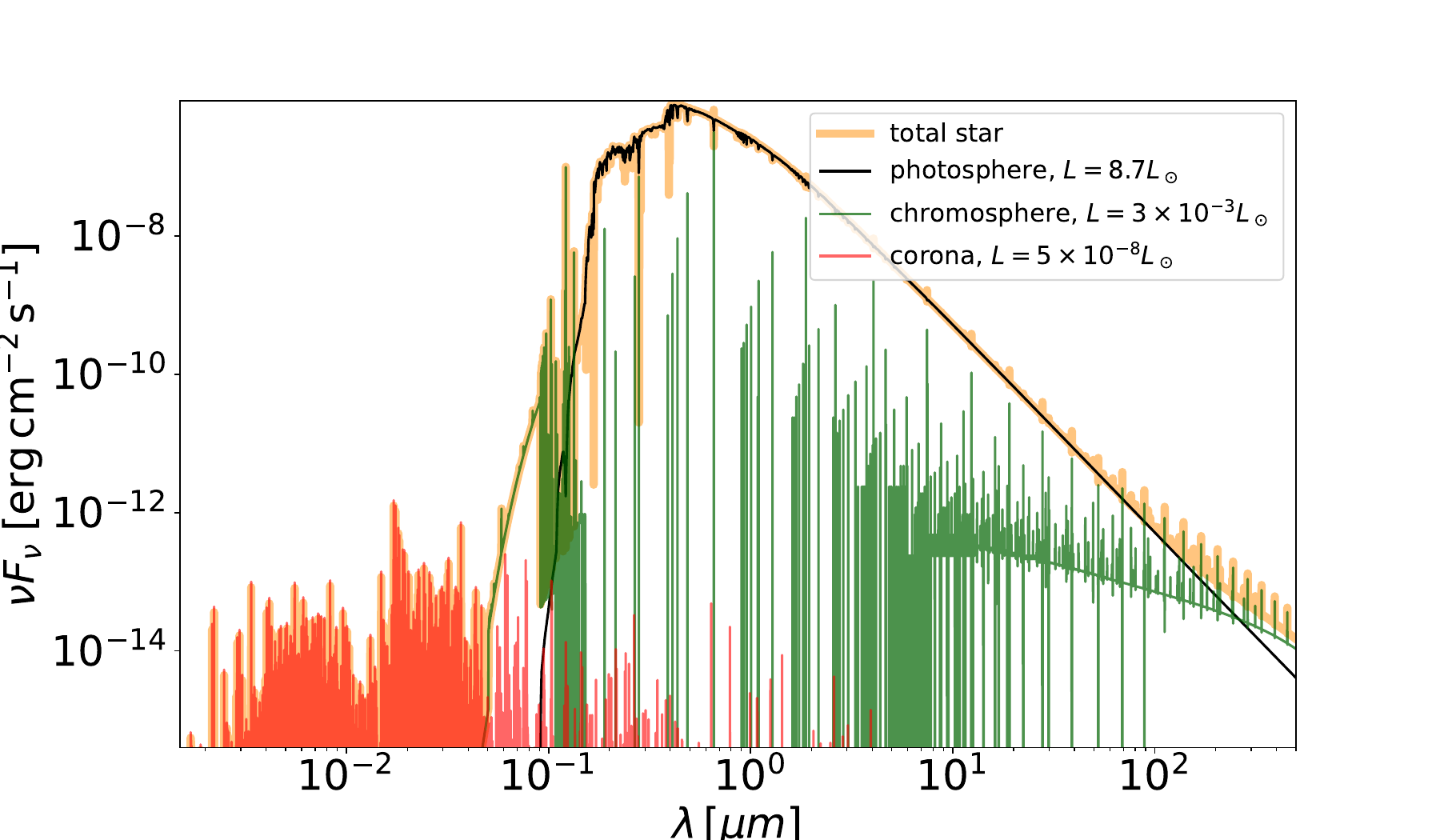}
\caption{Our modelled emissions from the stellar photosphere, chromosphere and corona. The chromosphere mostly emits in FUV, though  it also contributes in many line emissions. The corona is seen in X-rays.
}
\label{fig:3parts}
  \end{figure}

These are all highly suggestive of a hot chromosphere.  \citet{Bouret2002} 
constructed a simple model that is composed of a thin layer of gas ($N_{\mathrm{H}}$ $\sim 3\times 10^{20}$\,cm$^{-2}$) with a temperature that ranges from $10^4$ to $10^5\K$. They obtained a cooling luminosity $2.5\times 10^{-5}\,L_\odot$ in these lines alone.
We follow their lead but invoke an even simpler model, one where the column density is the same as theirs,
but where the temperature is constant. 
We increase the temperature from the photospheric value upwards, until the observed fluxes in the \ion{C}{2} and \ion{O}{6} lines can be reproduced.\footnote{\bpic\ is rapidly rotating with a rotational velocity of $v\sin i = 140$\kms. To emulate this effect in CLOUDY, we adopt a turbulence speed of 140\kms. This does not much affect our results.
}
We find $T \sim 1.3\times 10^4\K$. For this simple model, the 
radiative cooling luminosity, emitted mostly in the UV continuum, is $\sim 3\times 10^{-3}\,L_\odot$, comparable to that of the Solar chromosphere.
This energy is likely provided, as in the case of the Sun, from the mechanical heating of acoustic or Alfv\'en waves, excited below the photosphere.

\subsection{Corona model}


X-ray photons are also detected from \bpic\ \citep{Hempel2005, Gunther2012}. We again adopt a single one-zone model for the corona. The temperature is taken to be $1.1\times 10^6\K$, as determined by \citet{Gunther2012}. The total coronal luminosity is set to be $5\times 10^{-8} L_\odot$, to be compatible with the observed x-ray fluxes. This value is lower than the Solar coronal luminosity of $10^{-6}L_\odot$.

The coronal emission is too weak to affect the energy budget of the disk gas, but it does enhance the abundances of some highly ionized species (e.g., C$^{++}$, S$^{++}$). 

The combined stellar emission spectrum is shown in Fig.~\ref{fig:3parts}. The chromosphere and corona dominate over the stellar photosphere in wavelengths short-wards of $0.1\,\mu$m.

\begin{figure}
\centering
\hspace{-0.25in}
\includegraphics[width=0.49\textwidth]{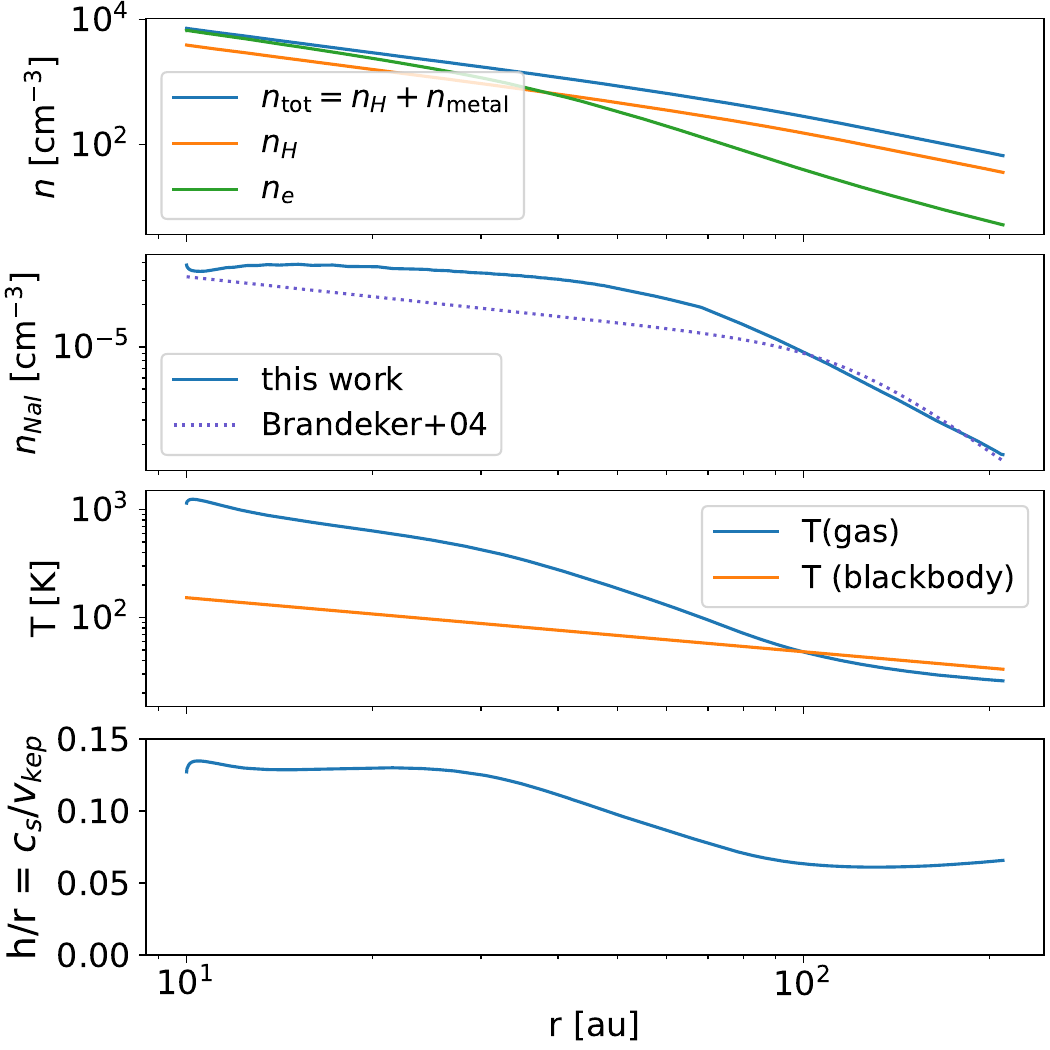}
\caption{Radial structure of our model gas disk. The top panel shows \uu{how number densities change with radius}: total (metal+hydrogen), hydrogen and electron. Most of the electrons are contributed by C/O. The second panel compares the Na I mid-plane number density, as calculated in our model and as empirically determined by \citet[their eq.~1]{Brandeker2004}. The third panel
displays the temperatures of gas (as determined by CLOUDY) and dust (black-body, zero albedo). The last panel is the vertical gas scale height for hydrostatic equilibrium.
}
\label{fig:structure}
  \end{figure}

\subsection{Accretion onto \bpic?}

It is surprising that \bpic, a A6V star (also a $\delta$-Scuti variable), should host a chromosphere and corona, structures that are usually associated with stars of later spectral types. 
Could the X-ray and UV emission be powered by gas accreting onto the star?

\uu{We present arguments against this explanation.} The line luminosities in the UV amount to a few $10^{-5}\,L_\odot$ \citep{Bouret2002}, and can be powered by an accretion at the rate of $\dot M \sim 6\times 10^{-13}\,M_\odot$\,yr$^{-1}$.  
This is already higher than a crude estimate from \citet[see, e.g.][]{King1992}. But it gets worse. To produce these lines, the hot gas requires a sufficient column (see above section). It will then also have to emit in the continuum, and according to our model, at a much higher luminosity of $3\times 10^{-3}\,L_\odot$. 
If accretion is to supply this, an unrealistically high accretion rate of $10^{-10}\,M_\odot$\,yr$^{-1}$ is required. 

The total mass of the \bpic\ gas disk is estimated to be $\sim 10^{25}\g$ (see below), so it could only supply the above high rate for $\sim 40$\,yr, i.e., a couple of dynamical time scales at 10\,au. Accreting disk gas can thus not be responsible for the UV emission. 



\section{Disk Model and Comparison with Observations}
\label{sec:diskmodel}

\subsection{Disk Model}

Here, we introduce a model for the gas disk that can 
reproduce the bulk of the observables. The model is kept deliberately as simple as possible, ignoring many complications (e.g., detailed radial profile, vertical distribution, azimuthal asymmetry, dust).

We assume the gas disk to extend from 10 to 200\,au, with a (vertically constant) number density of H that goes as
\begin{equation}
    n_\mathrm{H} = 220 {\cm^{-3}} \,
\left[\left({r\over{100\,{\rm au}}}\right)^{2.5} + \left({r\over{100\,{\rm au}}}\right)^{4.6}\right]^{-1/2}\, .
    \label{eq:nH}
\end{equation} 
This is similar in form but slightly different in parameters from that  adopted by \citet{Zagorovsky2010}. It  ensures that 
our derived \ion{Na}{1} profile closely emulates the observed one \citep{Brandeker2004,Nilsson2012}, as is shown in Fig.~\ref{fig:structure}.
The bulk of the gas lies at around 100\,au, consistent with the \ion{C}{2} study by \citet{Cataldi2014}. 
While our chosen outer cutoff is motivated by the observed extent of the gas dust disks, and does not strongly impact the results, 
the inner cutoff requires some explanations. 
Our choice of inner cut-off is motivated by the study of \citet{Brandeker2004}, where 
they observed \ion{Na}{1}  emission from 13 to 323\,au. 
In agreement, the kinematically resolved profile of the \ion{C}{2} 158\,$\mu$m line shows flux from only outside 10\,au \citep{Cataldi2014}. We also find best agreement with data when an inner cutoff is imposed at 10\,au \uu{(more below)}. Intriguingly, 10\,au is also the semi-major axis of the giant planet \bpic\,b.



Vertically, our model disk  extends a height of $H/r=0.2$ from the mid-plane (or covering a solid angle $\Omega = 4\pi \times 0.2$\,sr). This is \uu{chosen to be} somewhat thicker than the self-consistently calculated gas scale height, $H/r = c_s(T)/v_{\rm kep}$, which ranges from $0.05$ to $0.13$ (Fig.~\ref{fig:structure}),  \uu{so as to include most of the disk gas.}  

The composition of the disk is unusual. We assume it to be depleted of primordial hydrogen \uu{and helium. Any hydrogen is instead} supplied by sublimation from ices. The $N_\mathrm{H}/N_\mathrm{O}$ number ratio is taken to be $2$, reflecting H$_2$O. We assume there is no He in the disk, since it is unlikely to be retained by condensation or clathration. We then assume C, N, and Ar are in solar proportion to O \citep[solar composition from][]{Asplund2009}, while all other elements (see Fig.~\ref{fig:elements}, called `refractory' from now on) are reduced by a factor of $100$ from their solar values, again relative to O. \uu{The same
abundance applies throughout the gas disk.}

\uu{For the density profile in eq.~\refnew{eq:nH}, and our assumed abundance pattern,}
the total mass in this metal-rich disk amounts to $\sim 2\times 10^{25}\g$ ($\sim 0.3\,M_{\rm moon}$).  Out of this mass, only a very small fraction ($8\times 10^{22}\g$, or $0.3\%$) is contributed by the refractory elements. Moreover, the total mass of C is $\sim 6\times 10^{24}\g$
, consistent with that derived by \citet{Cataldi2018} using \ion{C}{1} and \ion{C}{2} lines. 

\begin{figure*}[t]
\centering
\hspace{-0.25in}
\includegraphics[width=0.99\textwidth]{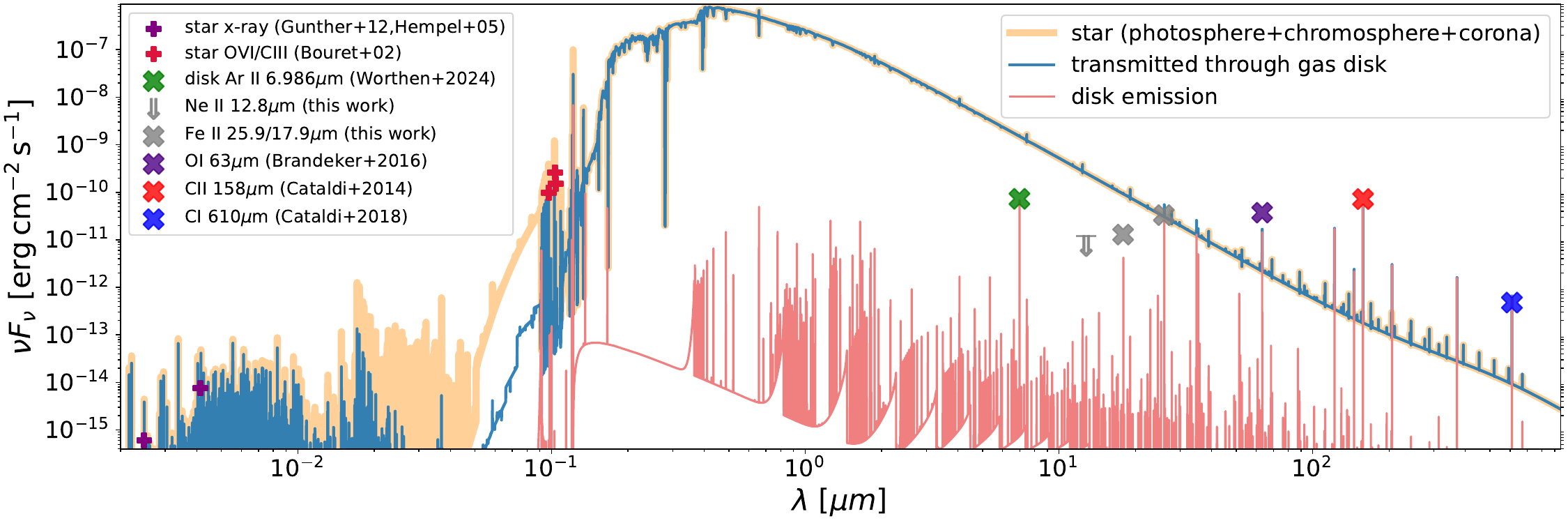}
\caption{Spectral energy distributions in our model, including the stellar spectrum (original, including all three emission components; transmitted: after passing through the disk) and that of the disk emission. Spectral resolution for the lines is taken to be  100\kms. 
The \ion{Ar}{2} line is the dominant gas coolant in MIR, explaining its detection by JWST. Fluxes in various lines agree reasonably well with the observed values (colored symbols). \uu{We do not include emission from the dust disk.}  
}
\label{fig:spectrum}
  \end{figure*}

  \begin{figure*}
\centering\includegraphics[width=0.99\textwidth]{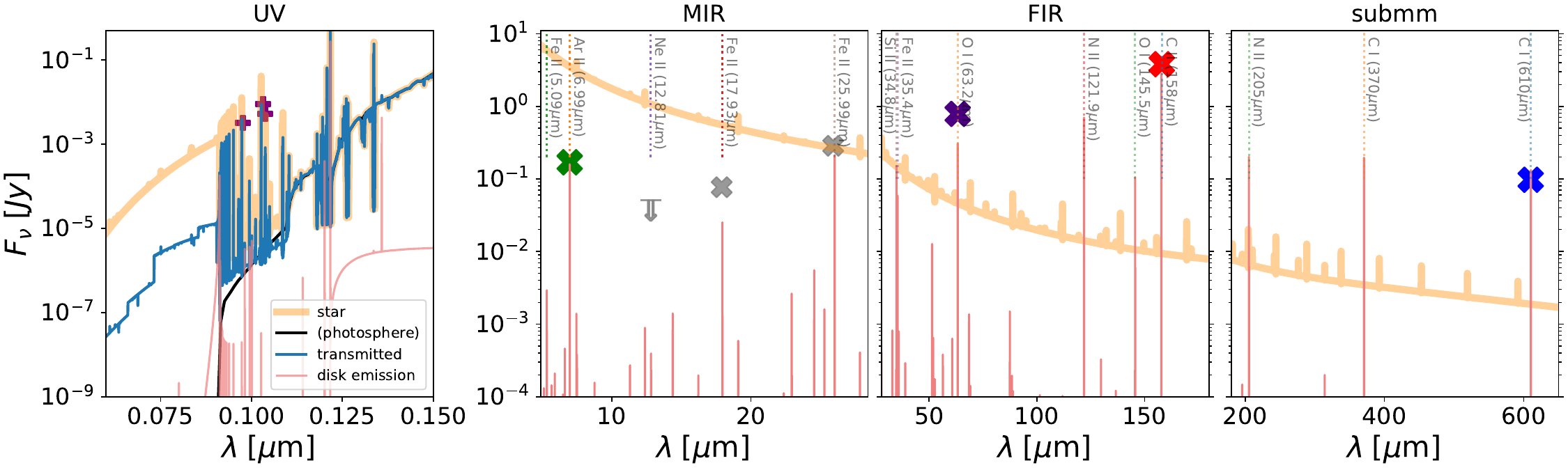}
\caption{A close-up view of Fig.~\ref{fig:spectrum} in four wave-bands. Same notations except the vertical axis is now flux\uu{, not $\nu F_\nu$. }
}
\label{fig:lines}
\end{figure*}

\begin{figure}
\centering
\includegraphics[width=0.45\textwidth]{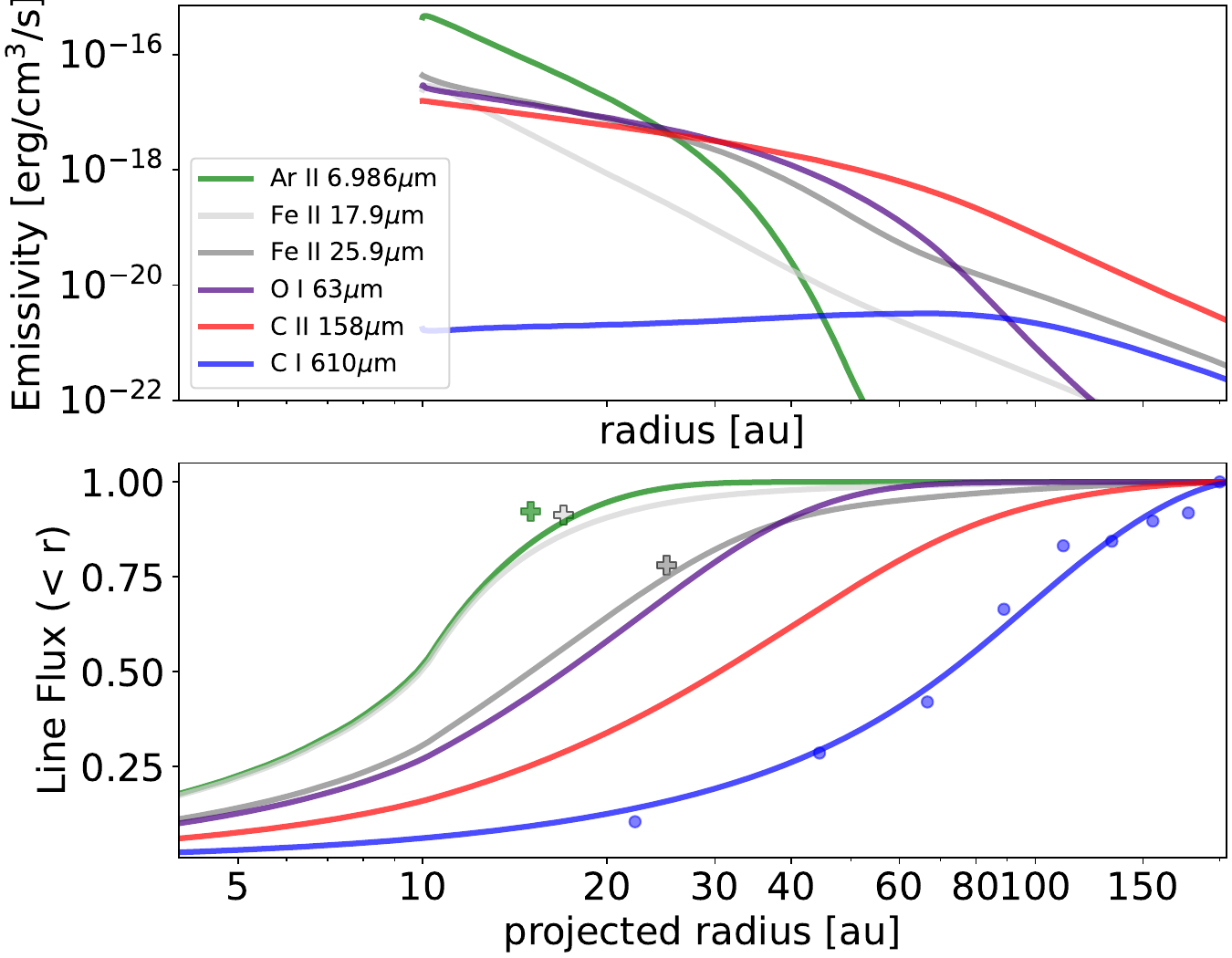}
\caption{Comparing radial profiles for the few important cooling lines. The top panel shows line emissivity as a function of radius, while 
the lower one the integrated line fluxes\uu{(assuming optically thin emission)} out to a given projected radius, with the total fluxes normalized to unity. \uu{Model indicates that the \ion{O}{1} 63\,$\mu$m line is moderately optically thick (Table \ref{tab:lines}).} The three crosses indicate measurements from JWST (Appendix \ref{sec:app1}), while the blue points are deduced from \citet{Cataldi2018}. There are good overall agreements between model and data.}
\label{fig:radialline}
  \end{figure}

\subsection{Comparisons\label{sec:comparisons}}

We use CLOUDY \citep{Ferland2013} to compute the thermal and ionization balances in the gas disk, under the irradiation of our model star. The spectral resolution in the CLOUDY simulations is set to\footnote{\yy{This
value merely sets the contrast ratio between the line and the continuum. The actual line-width used in CLOUDY for radiative transfer calculation is assumed to be thermal.}} $100$\kms.
In Fig.~\ref{fig:structure}, we show the resultant disk properties. Photo-ionization 
dominates disk heating and raises the disk temperature to over $1000\K$ near the inner boundary. This is dramatically hotter than those of previous models \citep{Zagorovsky2010,Kral2016}, and results from the high UV flux from our assumed chromosphere. The disk cools primarily through infrared and sub-mm metallic lines (including our main protagonist, the \ion{Ar}{2} line), as well as the Ly$\alpha$ line. 

The detailed radiation signature of such a disk is shown in
Fig.~\ref{fig:spectrum} (and a zoomed-in version in Fig.~\ref{fig:lines}). We display both 
the transmitted stellar spectrum (through an edge-on disk) and the emission spectrum from the disk itself. 

The transmitted stellar spectrum is consistent with the observed one. The edge-on disk absorbs most of the stellar UV, largely erasing the signature of the chromosphere except in some emission lines; the disk is transparent to the coronal X-ray. \uu{Our model largely reproduces the observed fluxes in UV lines and the X-ray bands. This is not surprising, as our stellar emission is tuned to do so.}

\uu{Line fluxes in the sub-mm/MIR are more revealing. CLOUDY predicts most of the cooling luminosity in this range to be in the form of C, O, Ar, and Fe lines. And by luck, the combination of ALMA, Herschel, and JWST capture them all. Moreover, 
we find that the model predictions agree with the observed values to within a factor of $3$ or better (also see Table~\ref{tab:lines}).\footnote{\uu{We under-predict the flux in the \ion{O}{1} 63\,$\mu$m line by about a factor of $2$. This line is moderately optically thick in the radial direction, so adding more oxygen will not increase the flux, but having a higher gas temperature (especially in the outskirt) will. This may indicate that a heating source is missing from the model, possibly that by dust.}} Since the line fluxes are affected by factors like gas density, gas temperature and ionization balances, it is remarkable that our simple model (with only  a handful free parameters) can reproduce these observations. The Fe lines, in particular, strongly support the over-abundances of the volatiles over the refractories. }

\uu{There are agreements beyond the emission line fluxes.}
For lines that are spatially resolved, we present a more fine-grained comparison in Fig.~\ref{fig:radialline}.  As one moves outwards in disk, CLOUDY predicts that the gas cooling is sequentially dominated by the \ion{Ar}{2} 6.986\,$\mu$m line, the \ion{O}{1} 63\,$\mu$m line, and 
lastly, the \ion{C}{2} 158\,$\mu$m and \ion{C}{1} 610\,$\mu$m line. The \ion{Ar}{2} line should be the most centrally concentrated because it requires a high UV flux. \uu{This agrees with the JWST data (analyzed in Appendix~\ref{sec:app1}) where  the majority of the \ion{Ar}{2} flux arises from inward of 20--30\,au. Moreover, if the disk extends to much inward of 10\,au, as opposed to truncates there,  there will be too much \ion{Ar}{2} flux from the unresolved region, compared to the observed case. This suggests that there is a gas hole inward of 10\,au.}
The two \ion{Fe}{2} lines are also similarly centrally concentrated, \uu{in both model and JWST observations (Appendix \ref{sec:app1}).} In contrast,  the \ion{C}{1} line traces cold neutral gas \uu{that only exists in the outskirt. It is thus} the most spatially extended, with emissivity peaking near 100\,au, again in good agreement with results from \citet{Cataldi2014}. 
In fact, the apparent \ion{C}{1} emission cavity (extending out to 50\,au) as discussed by \citet{Cataldi2018} is not due to a genuine cavity but to the low emissivity of \ion{C}{1} in the inner hot region. 


\begin{figure*}
\centering
\includegraphics[width=0.99\textwidth]{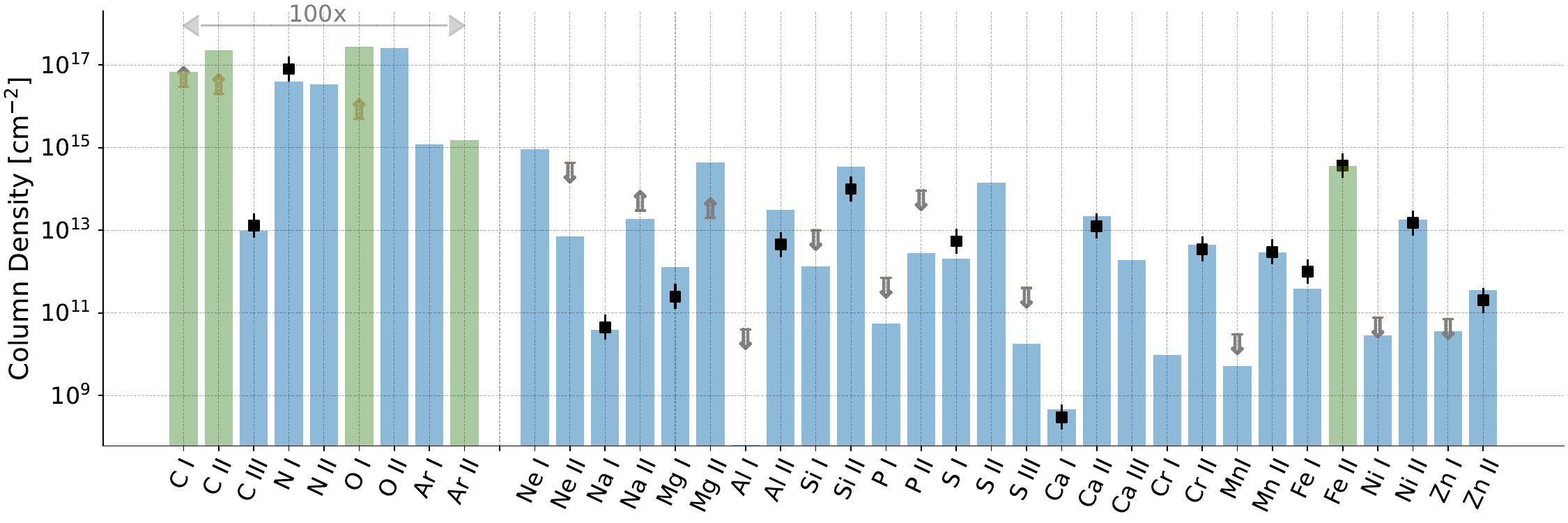}
\caption{A comparison of the column densities of various elements and their ionized varieties, between those observed \uu{from absorption line studies} (black symbols, from Table \ref{tab:elements}, error bars indicate a factor of $2$ uncertainty, gray arrows indicate upper or lower limits)  and those calculated by CLOUDY (bars). \uu{Green bars indicate column densities for the species where emission line fluxes are measured and are successfully reproduced by our model  (Fig. \ref{fig:spectrum}).}
The left group (C, N, O, and Ar) are assumed to be enriched by $100\times$ over the other elements. 
There is an overall good agreement between the model and available constraints, \uu{across 8 orders of magnitude in column densities}. }
\label{fig:elements}
  \end{figure*}

We now turn to a comparison with absorption line studies. Fig.~\ref{fig:elements} displays our predicted column densities for various atoms and ions, together with data from literature (Table~\ref{tab:elements}). It is remarkable to observe that, while the data contain some 30 species and span some 8 orders of magnitude in values (including upper/lower limits), our simple model is able to \uu{reproduce the observations}, mostly within a factor of two or better.\footnote{\uu{The most glaring exception to this overall agreement is \ion{Al}{2} (the dominant ionization state of Al),  where we over-predict its column density by a factor of $7$. The significance and the meaning of this difference  is not clear to us. }} In particular, while the original \citetalias{Wilson2019} determination of N$^0$ would have indicated a depletion of N, which would be surprising astro-chemically, our new determination (Appendix~\ref{sec:N_data}) brings N comfortably inline with other volatile species. 

\begin{figure*}
\centering
\includegraphics[width=0.95\textwidth]{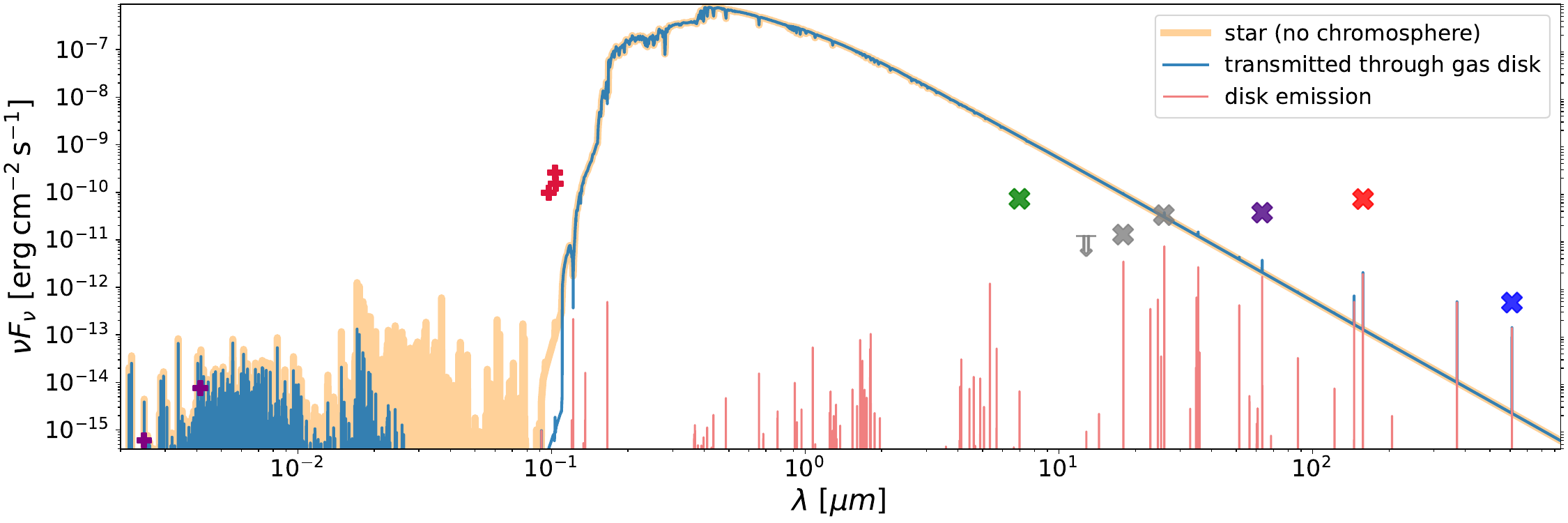}
\caption{Same as Figs. \ref{fig:spectrum} but for a model where the chromosphere is absent. The cooling line fluxes are all much reduced, most notably the \ion{Ar}{2} line, as the gas is less photo-ionized and is cooler.
}
\label{fig:corona}
  \end{figure*}

Lastly, we wish to emphasize the importance of the stellar chromosphere in these successes. 
If instead \bpic\ does not possess a chromosphere, the resultant gas emission  is shown in Fig.~\ref{fig:corona}. 
The \ion{Ar}{2} line is nearly absent, as there is not enough flux to ionize Ar (first ionization potential 15.6\,eV). 
In general, photo-ionization heating is much reduced,\footnote{If so, photo-electric heating from the dust grains will dominate gas heating \citep{Zagorovsky2010}.} 
leading to lower gas temperatures and much diminished cooling line fluxes. We also fail to reproduce the abundances of many ionized species. 





\section{Gas production in Debris Disk}
\label{sec:discussions}

Our deduced abundances, a Solar-pattern except for a factor of 100 enhancement for C, N, O, and Ar, place strong constraints on the capture process in the proto-planetary disk (next section), as well as on the release process in the debris disk (this section). 

\subsection{Preamble}

Gas is most likely released from \uu{small} dust grains.\footnote{\yy{''While larger bodies may contain a larger reservoir of volatile
gases, their smaller surface areas likely make them irrelevant. After
their fresh surfaces are out-gassed down to a small excavation depth, volatiles diffuse out
slowly through the surface substrates. However, see \citet{Bonsor2023} where they explored the effects of very large excavation depths.'}} Here, we present grain properties pertinent to out-gassing: temperature, dust mass and collision timescale. We  assume that most of the dust surface area (for scattering and thermal emission) is provided by the smallest grains stable to radiation pressure, $s\sim 5\,\mu$m \uu{\citep{Artymowicz1988}}.

The blackbody temperature scales with radius as (with albedo $A_b$) 
\begin{equation}
T_{\rm bb} = 47\K \,  (1-A_b)^{-1/4}\, \left({r\over{100\,{\rm\,au}}}\right)^{-1/2}
\, .
\label{eq:Tbb}
\end{equation}
A 5\,$\mu$m grain is slightly warmer than this, because it is inefficient in radiating at thermal wavelengths. We ignore this difference.

The dust mass can be estimated from the dust luminosity \citep[$L_{\rm excess}/L_* \sim 2\times 10^{-3}$, e.g.,][]{Nilsson2009}, which equals the fractional area covered by the grains, $N_{\rm grain} \pi s^2/(4\pi r^2)$. Taking a grain 
bulk density of $\rho_{\rm grain} = 1\g/\cm^{3}$ and placing all grains at $\sim 100$\,au, we arrive at a small dust mass of $M_{\rm small-dust} = (4/3) \pi s^3 \rho_{\rm grain} N_{\rm grain} \sim 2\times 10^{25}\g$, nearly the same as the total gas mass we have inferred above. 
This estimate does not include contribution from larger bodies.

To estimate the collision timescale, we assume that the debris disk is an annulus of radius $r\sim 100$\,au, of width $\Delta r \sim r$, and a vertical scale height $H$. The radial optical depth of dust grains is $\tau_r \sim (L_{\rm excess}/L_*)/(2H/r)$, and 
the vertical optical depth is,
\begin{equation}    
\tau_{\perp} \sim  \left({{2H}\over{r}}\right) \, \tau_r \sim \left({{L_{\rm excess}}\over{L_*}}\right) \sim 2\times 10^{-3}\, . 
\label{eq:tauperp}
\end{equation}
This is also the chance for grains to collide per half an orbit, yielding a mean collision time of 
\begin{equation}
    t_{\rm coll} \sim {1\over {2\tau_\perp}}\, P_{\rm orb} \sim  \, 2\times 10^{5} \yr \, \left({r\over{100 \rm\,au}}\right)^{3/2}.
    \label{eq:tcoll}
\end{equation}
In comparison,  the estimated stellar age is $\sim 20$\,Myr.

\subsection{Gas Release}
\label{subsec:production}

To ensure a uniform enrichment of the volatiles in the gas, it is best if they are released from the grains via the same process.
The hyper-volatiles (in the form of CO, N$_2$ and Ar) are efficiently thermally desorbed when temperature of the matrix (water ice) reaches $T \geq [35,40]\K$ \citep{Simon2023}.  
This is satisfied in most of the \bpic\ disk (eq.~\ref{eq:Tbb}, Fig.~\ref{fig:structure}). In contrast, water ice cannot thermally desorb  (sublimate) when $T \leq 100\K$,\footnote{The thermal desorption rate \citep[see, e.g.,][]{Fraser2001}.
\begin{equation}
{{dN}\over {dt}} = - A \exp \left(-{{E_d}\over{k_B T}}\right)\,
\label{eq:tdesorb}
\end{equation}
where experimentally determined values for water ice are $A \sim 10^{30}$\,molecules\,cm$^{-2}$\,s$^{-1}$, and $E_d/k_B \approx 5800\K$. } i.e., over much of the observed disk \citep[also see][]{Grigorieva2007}.

In the high-UV environment of \bpic, water can instead be released by photo-desorption (a.k.a.\ photo-sputtering), a non-thermal process of out-gassing induced by UV photons. This was first proposed by \citet{Artyomowicz1997} and discussed in \citet{Chen2007,Grigorieva2007}. 

Theoretical calculations \citep{Andersoon2008} found that ice has a significant absorption cross section only in the 7.5 -- 9.5\,eV range (or 1300--1653\,\AA). Photons within this range are predicted and measured to have a desorption efficiency of  
 $Y_{\mathrm{H_2O}}=$ 1 molecule/1000 UV photons.
\citep{Andersoon2008,Westley1995,Oberg2009,Cruz2018}. Such a low value of $Y_{\mathrm{H_2O}}$ likely results from the layered structure -- while UV photons can penetrate down to of order $\mu$m in depth,
only those molecules liberated in the first handful of mono-layers
can easily escape \citep{Andersoon2008,Oberg2009}. 

In \bpic, such UV photons are mostly produced in the photosphere (and not in the chromosphere) and account for $\sim 10^{-3}$ of the total stellar flux.
We find that a mono-layer of water ice (with an assumed molecular radius 3\,\AA) is lost after
\begin{eqnarray}
t_{\rm photo, mono} & \approx & \left[{Y_{\mathrm{H_2O}}
    \times \left({L_{\rm UV}}\over{h\nu 4\pi r^2}\right) \times \pi (3\,\mathrm{\AA})^2}\right]^{-1}\nonumber \\
& \sim & 0.1\,{\yr} \left({r\over{100\,{\rm au}}}\right)^{2}\, \, \left({{L_{\rm UV}}\over{10^{-3} L_*}}\right)^{-1} \, \left({{L_*}\over{8.7 L_\odot}}\right)^{-1}
\,,
    \label{eq:tphoto}
\end{eqnarray}
\yy{where $L_*$ is \bpic's bolometric luminosity and is taken to be  $8.7L_\odot$.}
It takes a mere (5\,$\mu$m/3\,\AA)$\times t_{\rm photo,mono} \sim 1700$\yr\ to evaporate a 5\,$\mu$m grain, if it is made up of pure ice \citep[also see][]{Grigorieva2007}.\footnote{\citet{Grigorieva2007} adopted different low energy cutoff and different absorption coefficient from us, leading to a factor of $2$ difference in the desorption rate.} 
This is much faster than it can be replenished by grain collisions (eq.~\ref{eq:tcoll}). Moreover, given that $M_{\rm small-dust} \sim M_{\rm gas}$, evaporation at this high rate would build up the entire gas disk in a few thousand years.\footnote{Such a rate, $\sim 10^{22}\g/\yr$, corresponds to the upper limit set by \citet{Cavallius2019}, based on a non-detection of water vapour and a water  photo-dissociation lifetime of 3.5\,d at 80\,au \citep{Cataldi2014}. } Both seem unreasonable. 
So the small grains observed in \bpic\ cannot be predominately icy, a conclusion also reached by \citet{Grigorieva2007}.

Realistic planetesimals are likely `dirty' -- aggregates of ice (`wet') and refractory (`dry') clusters that result from condensation in the proto-planetary disk. The dry part, even at a few mono-layers thick, may stall volatile loss, for the same reason that $Y_{H_2O}$ is small (see above). Subsequent out-gassing is possible only after fresh icy surfaces are exposed by collision, or if the volatiles escape through narrow tunnels and surface cracks. Both may reduce the rate of water photo-desorption substantially.

Such a scenario of slow desorption may be tested by studying grain mineralogy. We expect 
the small grains in \bpic\ to retain most of their refractories and some of their ices. The ices are buried below a thin layer of substrate but can resonantly scatter/emit photons with wavelengths longer than a few microns \citep[for a thorough investigation, see][]{Kim2019}. As a result, the dust disk should appear bright in both the $10\,\mu$m silicate feature \citep[as is indeed observed by][]{Chen2007,Lu2022}, and the $3\,\mu$m ice feature.  So far, the ice signature has not been found in debris disks \citep{Grigorieva2007,Rebollido2024}, but we predict that JWST should detect it. 

In summary, we suggest that water ice is photo-desorbed from small grains at a much reduced rate than theoretically possible, due to the `dirty' grain composition. The hyper-volatiles are released concurrently with water, explaining the near-primordial C, N, O, and Ar abundance pattern. 

What about the refractory elements (including P and S)? They can also be released by photo-desorption. An early attempt by \citet{Chen2007} arrived at a Na desorption rate of 
$\sim 10^{18}\g\yr^{-1}$ in the \bpic\ disk, based on limited lab experiments. But currently, we have little constraint on the desorption yield, and have little understanding on  how the yield depends on photon energy and on atomic species.
Given the near-solar abundance pattern for the refractories in the gas disk, as well as their co-spatial  distribution with the volatiles (see, e.g., Fig.~\ref{fig:structure}), 
we instead suggest that they are more likely the by-product of volatile out-gassing. But the actual process is unknown. \uu{To account for their under-abundances, we speculate that refractory nuggets may survive the collisional cascade until they reach the radiative blow-out size. Their removal in this form may account for their depletion. If so,  we expect the $\beta$ meteorites (small grains radiatively accelerated to elliptical orbits)  to be more silicate-rich.}

\subsection{Alternative models}

\uu{Alternatively, gas may evaporate out of grains during grain collisions.}
\citet{Czechowski2007} pointed out that collisions between bound grain (on largely circular orbits) and $\beta$ meteoroids can reach speeds high enough for vaporization \citep[e.g.,  $6.5\km/\s$ for ice and $\sim 19$\kms\ for silicates,][]{Tielens1994}. 
Based on a model of grain vaporization and fragmentation from \citet{Tielens1994}, they calculated a gas production rate of $\sim 1.6\times 10^{19}\g\yr^{-1}$ for silicate in the \bpic\ disk, about 40$\times$ \uu{lower} than that for water.
\uu{At face value,} these rates appear competitive against those from photo-desorption \uu{and can give rise to a high O/Si ratio. However, these calculations assume that grains are hit by grains of the same composition. This is un-realistic as the grains co-exist in the same disk. In fact, the volatiles may be well hidden under a surface substrate of silicates, as a result of their rapid photo-desorption (see above). Moreover, different refractory elements will likely have different vaporization energy, making it hard to explain their uniform abundances.}



\citet{Czechowski2007} also discussed sputtering by energetic particles in the stellar wind \citep{Tielens1994}. The strength of stellar wind from \bpic is unknown. Moreover, the sputtering yield depends on the elements strongly \citep[see Fig.~10 of][]{Tielens1994}. It is therefore hard to foresee how this will generate the observed enrichment of the volatiles.



\subsection{Caveats and Predictions}

\uu{Having offered our criticisms for these alternative scenario, we return to examine caveats for our} favored model, photo-desorption.  In addition to those discussed in \S \ref{subsec:production},  the following caveats are present:
\begin{itemize}
\item there exists the possibility that the volatile gas can be re-absorbed onto grains, a process invoked to explain the HD\,32297 disk \citep{Cataldi2020}. The timescale for this (due to gas-grain collision) is competitive with photo-desorption and grain-grain collision timescales;
\item the radial density profile of the gas disk \uu{remains puzzling. While the highest dust concentration is near 100\,au \citep{Artyomowicz1997}, the gas disk has a density that rises inward towards 10\,au, as evidenced by both the \ion{Na}{1} profile \citep{Brandeker2004} and the bright \ion{Ar}{2}, \ion{Fe}{2} lines seen by JWST \citep[][and Appendix \ref{sec:app1}]{Worthen2024}. The ionization state of the gas disk is likely high enough for a magneto-rotational instability. So magnetic stress could have led to radial diffusion \citep{Kral2016}. But such a picture is in conflict with} the strong azimuthal asymmetry seen in \ion{C}{1} \citep{Cataldi2018}. \uu{The absence of gas inwards of 10\,au is also hard to explain.}\footnote{Though the giant planet, \bpic\,b, orbiting at $\sim 10$\,au, may complicate this argument.}




\item \uu{along the same vein,} it remains an unexplained coincidence that the current gas disk is comparable in mass to that in the current generation of small grains. This is a surprise if the gas is accumulated over many collision times. 

\end{itemize}

\uu{But if the gas disk is predominantly produced by photo-desorption of dust grains, one expects a couple observable consequences:}
\begin{itemize}
\item \uu{the bound dust grains should be a mixture of ice and silicate material, while the un-bound grains should be mostly silicate, as discussed in \S \ref{subsec:production}.}

\item \uu{debris gas disks should preferentially orbit early-type stars. The near-UV (7--10\,eV) luminosity in lower-mass stars (FGKM) are too low to desorb much gas out of the dust grains.}
\end{itemize}

\section{Elemental Abundances of Planet-forming Material}
\label{sec:abundances}

Our simple model have successfully reproduced a wide range of observables in the \bpic\ disk. Here, we capitalize on this success to answer a second question: what could the abundance pattern in the \bpic\ gas disk inform us about the planet formation process?

In this work, we find that Ar and N are just as enriched as C and O. Such a uniform enrichment, despite disparate physical and chemical properties, is best explained if  almost all C, N, O, and Ar in the proto-planetary disk are retained, \uu{either through direct condensation, or more} probably, by entrapment on (water)-ices. Ar and N place the strongest constraint on the capture process as they are the most volatile. 
%


Second, \uu{the ratio of O/C can reveal the nature of volatile capture. For direct condensation, one expects a primordial ratio of O/C; while a complete entrapping of C, N, and Ar by water requires a large amount of O, } 
\uu{almost 4 times} as much O as the solar \uu{O/C ratio suggests} (see \S \ref{subsec:intro2}).
%
\uu{In \bpic\ disk, while \ion{C}{1} and \ion{C}{2} are well measured from emission lines, the O abundance is more uncertain.\footnote{The UV absorption lines of these elements are all heavily saturated \citep{Roberge2000}.} We are embattled by the fact that the \ion{O}{1} 63$\,\mu$m line is both optically thick (Table \ref{tab:lines}) and not spatially resolved \citep{Brandeker2016}.} 
\uu{We only have a weak constraint on the O/C ratio, $N_\mathrm{O}/N_\mathrm{C} \sim (N_\mathrm{O}/N_\mathrm{C})_\odot$, within an order unity.\footnote{We have experimented with $N_\mathrm{O}/N_\mathrm{C} = 4 \times (N_\mathrm{O}/N_\mathrm{C})_\odot$. It  causes order of unity variations in the emission line fluxes. We cannot exclude it based on this crude model. }
So while how volatiles are captured is a highly important question, we  cannot at the moment resolve the issue.}

\begin{figure*}
\centering
\includegraphics[width=0.99\textwidth]{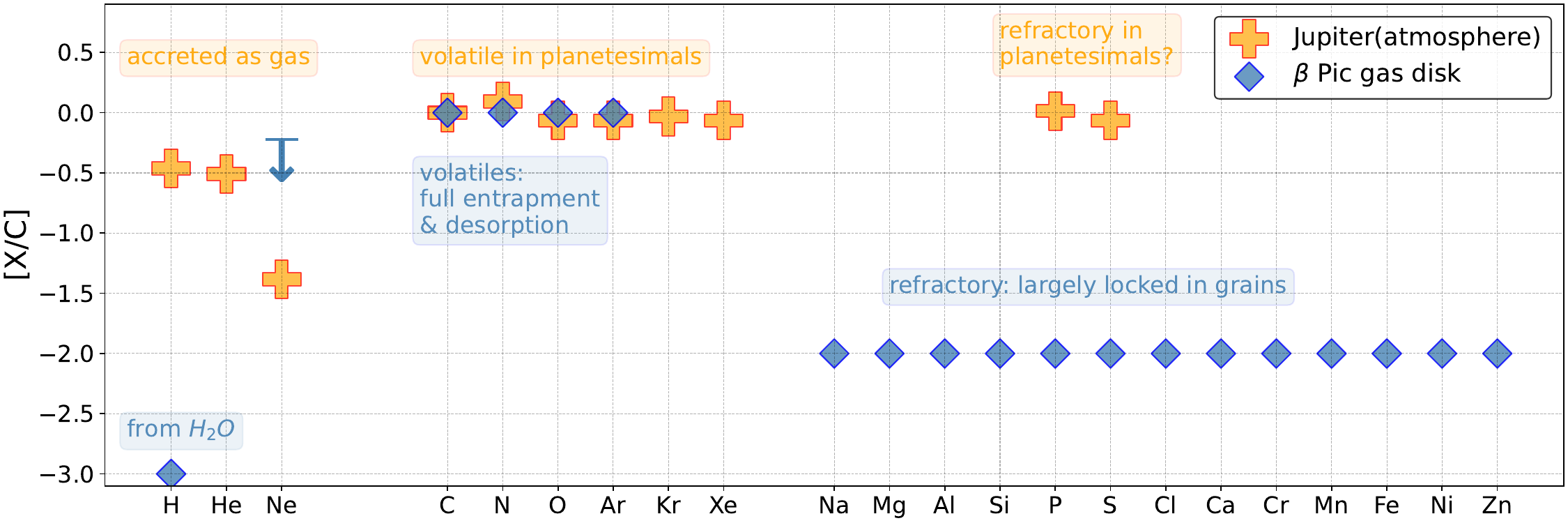}
\caption{Comparison of abundance patterns in \bpic\ gas  disk (blue) and in Jupiter's atmosphere (orange). All abundances are measured relative to C and normalized by the solar values. 
All error bars are suppressed for clarity. 
For \bpic, Ne is marked by its upper limit.
Text-boxes suggest possible pathways for these elements.
}
\label{fig:backtous}
  \end{figure*}

\uu{We now attempt to draw some insights by comparing against the abundance pattern in  Jupiter's atmosphere} (Fig.~\ref{fig:backtous}). 
For the latter, we adopt values compiled by \citet{Atreya1999} and updated in
\citet{depater2023}.
There are clear agreements and disagreements, all deserving some thoughts. In the following, 
we discuss the comparison, element by element:

\begin{itemize}

\item H and He:  current constraints on H and He in the \bpic\ disk are weak. We assume that H is not primordial and is only derived from H$_2$O dissociation, and that He is not entrapped by planetesimals and is absent. \uu{Jupiter, on the other hand, contains a massive amount of H/He that is accreted from the disk gas.}

\item Ne: similar to He, Ne has a small size and is difficult to \uu{condense or to} entrap into solids \citep{Lunine1985}.
So we do not expect to see Ne in the \bpic\ disk. Currently, only a weak upper limit on Ne can be set using the \ion{Ne}{2} 
$12.81\,\mu$m line (Table \ref{tab:lines}), 
\uu{corresponding to a Ne abundance that is at most as enriched as C, N, O, or Ar. }
Interestingly, Ne is also not enriched in Jupiter's atmosphere, \uu{consistent with it being brought in together with the H and He gas.}

Any detectable amount of Ne in \bpic\ would pose a strong challenge to the idea of volatile capture.
The best way to hunt for Ne will likely be the \ion{Ne}{1} absorption lines in the optical (e.g., 585, 640\,nm), since Ne
should mostly be in the form of \ion{Ne}{1} due to the latter's high ionization potential  (21.56\,eV). 
But these are transitions from excited states (not ground state) and may be challenging to detect. 

\item C and O: \uu{these are enriched in both  Jupiter and \bpic. As discussed, the ratio $N_{\mathrm{O}}/N_{\mathrm{C}}$ is of major interest, as it can arbitrate between direct condensation and water entrapment. This ratio is consistent with the solar value in both environments, and each with too large an uncertainty to distinguish between the two scenarios. }


\item N: the main carrier of $N$ in proto-planetary disks in $N_2$. \uu{N is enriched in both \bpic\ and Jupiter, again suggesting a common origin.} Given its property, we expect N to be enhanced wherever Ar is. 
Our new determination of $N^{0}$ supports this expectation (Appendix \S \ref{sec:N_data}).  For further support, we suggest absorption line studies of  N$^{0}$ or N$^{+}$ (they should be at roughly equal proportion).  
%
CLOUDY also predicts strong \ion{N}{1} 121.9\,$\mu$m and \ion{N}{1} 205\,$\mu$m lines (Fig.~\ref{fig:lines}).  


\item Ar, Kr, and Xe:
 known to be enriched uniformly (by $\sim 3\times$) in Jupiter's atmosphere, we expect Kr and Xe to be as enriched as Ar in the \bpic\ disk. \uu{Ar enrichment places the strongest constraint on the process of volatile capture.}

\item S and P: these are two interesting species. We do not find them to be enriched like the volatiles (though the constraint on P is weaker), suggesting that, if they are incorporated into planetesimals, they should be  in refractory forms (e.g., FeS), not volatile forms (e.g., H$_2$S). This is consistent with the fact that S and P are heavily depleted in the ISM and in cold disks \citep{Kama2019}.


This then begs the question of why Jupiter's envelope is enriched in S and P (by about a factor of 3). It is possible that these are released by bombarding planetesimals as they sublimate in Jupiter's atmosphere, \uu{even if they were initially in refractory forms.}  If so, we predict that all heavy elements (other than C, N, O, Ar, S, and P) should be similarly enriched in Jupiter's atmosphere.\footnote{\uu{Such a uniform enrichment requires a total solid mass of $\sim 12 M_\oplus$ in Jupiter's atmosphere ($ M_J \approx 300 M_\oplus$). It reduces to $8 M_\oplus$ if  only C/N/O/Ar are enriched.}}

\end{itemize}

These discussions make clear that, despite the different enrichment abundances, the planetesimals that out-gas in the \bpic\ disk \uu{may be} of \uu{a} similar composition to those that pollute the atmosphere of Jupiter. 

\section{Conclusion}

We construct a radiative transfer model that successfully explains an array of observables in the \bpic\ gas disk. \uu{These include column densities of a few dozen elements, fluxes in a handful of MIR, FIR, and sub-mm emission lines, and the radial profiles of these lines.} There are two key elements in this model. 
First, the host star needs to harbor a hot chromosphere, allowing it to ionize Ar and other elements to their observed states. This may be surprising for an A-type star, but is supported by many other lines of evidence.
Second, all elements follow roughly the solar pattern, except for an under-abundance of H (and possibly He), and an over-abundance of a uniform factor of $\sim 100$ for  the volatile elements (C, N, O, and Ar).

\uu{Uniform capture of these volatiles, especially the hyper-volatiles (N and Ar), into solids (parent planetesimals for the dust grains), requires a very cold temperature. In the case of entrapment by water, it also requires an over-abundance of water, with $N_\mathrm{O}/N_\mathrm{C} > 6.6$, almost 4 times that of the solar ratio. Unfortunately, we could not accurately determine the O/C ratio, mostly because the \ion{O}{1} line is optically thick, and partly because our model has inherent order-unity uncertainties. Our predicament in the case of \bpic\ disk mirrors that in the case of Jupiter's atmosphere.}

\uu{While we are encouraged by the success of the current model to reproduce a wide range of observations, it remains crude. It is essentially a 1-D radiative transfer model. Many more  nuances need to be taken into account: vertical height dependence, azimuthal asymmetries, dust contribution...  It is hoped that with a  more sophisticated model, one can place stronger constraints on some of the most interesting values, like the C/O ratio, the presence of Ne and other noble gases, and the abundances of S and P. }

\uu{We have also considered how the elements are out-gassed from the dust disk.}
The current gas mass in the \bpic\ disk is $\sim 2\times 10^{25}\g$ (or $\sim 1/3$ lunar mass), similar to that in the current generation of small grains ($\sim 5\,\mu$m).
Among this, water ices  \uu{and other volatiles} are \uu{easily} desorbed off the grains by the stellar UV fluxes.
The \uu{actual} photo-desorption rate, constrained by multiple lines of evidences, has to be much slower than that for pure-ice. So we argue that the grains are likely `dirty' aggregates, made up of icy and silicate clusters and where desorption is slowed down by the silicate (or graphite) surfaces. \uu{The volatiles are only released when fresh surfaces are exposed to starlight by collisions.} This could be tested by grain mineralogy studies, by targeting the water-ice features. Understanding the microscopic structure of planetesimals can illuminate their formation process.

The refractory elements, on the other hand, may also be released in the same process, albeit at a rate reduced by a factor of $100$. 
\uu{One possible way this occurs is, while volatiles are easily out-gassed from the micron grains, silicate nuggets may persist to sub-micron sizes and are subsequently blown-away by radiation pressure. One then expects the $\beta$-meteorites to be heavily silicate rich and ice poor.}

\bigskip

The trace amount of gas mass  in \bpic\ would have remained quite invisible, if not for the UV photons from the star: UV photons from the stellar photosphere desorb volatiles off the dust grains; then UV photons from the stellar chromosphere photo-ionize and heat the released atoms, making them visible to us in a multitude of emission and absorption lines.
And by observing these signatures, we can unwind the clock to study an earlier phase, when the planetesimals form out of proto-planetary disks. 


\bigskip
\medskip
We thank the organizers of the Dust Devils workshop  where this project was initiated. \uu{We are also grateful to an anonymous referee for a careful and thoughtful report, which prompts us to re-consider the importance of direct condensation.}
We acknowledge helpful conversation with Peter Martin, \uu{Quentin Kral and P.~A.\ Str\o m (n\'e Wilson).} YW's research is supported by NSERC.
KW and CC's work are supported by the National Aeronautics and Space Administration under grant No. 80NSSC22K1752 issued through the Mission Directorate. AB acknowledges support from the Swedish National Space Agency.


\bibliographystyle{aasjournal}
\bibliography{rings}{}

\appendix

\section{New measurements using JWST/MIRI MRS data}
\label{sec:app1}

\begin{table*}[]
\vskip0.2in
    \centering
    \begin{tabular}{{l|c|c|c|c}}
    \hline
        Line & Distance (au) & Resolved Flux (erg s$^{-1}$cm$^{-2}$) & Total Line Flux (erg s$^{-1}$cm$^{-2}$)& Resolved Flux/Total Line Flux\\
        \hline
         Ar II & 15--20 & 2.0$\pm$0.4$\times10^{-15}$& 2.6$\pm$0.1$\times10^{-14}$&0.08$\pm$0.02 \\
        Fe II 17.94\,$\mu$m & 17--22& 4.1$\pm$0.8$\times10^{-16}$& 4.7$\pm$0.5$\times10^{-15}$ &0.09$\pm$0.02 \\
        Fe II 25.99\,$\mu$m & 25--35& 3.1$\pm$0.6$\times10^{-15}$& 1.4$\pm$0.1$\times10^{-14}$&0.22$\pm$0.05\\
        \hline
   
    \end{tabular}
    \caption{Spatially resolved line fluxes from the MIRI MRS data. The distance column shows the projected distance from the star covered by the extraction aperture. This distance corresponds to the 3$\sigma$ contours shown in Figure \ref{fig:FeII_profiles}. The right most column shows the ratio of the resolved line flux to the total line flux. }
\label{tab:resolved_flx}
\end{table*}

We re-reduced the JWST MIRI MRS data of $\beta$ Pictoris that was presented in \citet{Worthen2024}\footnote{This work is based [in part] on observations made with the NASA/ESA/CSA James Webb Space Telescope. The data were obtained from the Mikulski Archive for Space Telescopes at the Space Telescope Science Institute, which is operated by the Association of Universities for Research in Astronomy, Inc., under NASA contract NAS 5-03127 for JWST. These observations are associated with program 1294. The specific observations analyzed can be accessed via\dataset[DOI]{https://archive.stsci.edu/doi/resolve/resolve.html?doi=10.17909/7xb7-hh14}. }
with a newer version of the JWST pipeline (version 1.14.0, Calibrated Reference Data System context ``jwst$\_$1223.pmap") and 
detect two new 
\ion{Fe}{2} lines. For a description of the observational parameters and setup, see \citet{Worthen2024}. We processed the data using the exact same pipeline steps as \cite{Worthen2024}, but then used the pipeline to extract the spectrum of the unresolved point source of \bpic\ with an extraction aperture of 1.5$\times$ the PSF FWHM at each wavelength. Because we are only interested in the emission lines, we removed the continuum from the spectrum by subtracting a B-spline fitted to the spectral points outside of each of the emission lines. From this newly reduced spectrum, we detect \ion{Fe}{2} emission lines at 17.936 and 25.988\,$\mu$m and we also place an upper limit on \ion{Ne}{2} at 12.838\,$\mu$m. The two \ion{Fe}{2} lines and the region of the spectrum containing the wavelength of the \ion{Ne}{2} line are shown in Fig.~\ref{fig:spectrum}.  

We fitted the two \ion{Fe}{2} lines with Gaussian profiles and then integrated the best-fit Gaussians to calculate the line flux of the spatially unresolved component of the line flux. The line fluxes are listed in Table~\ref{tab:lines}. We calculated a 3$\sigma$ Ne II line flux upper limit by computing the standard deviation (1$\sigma$) of the 10 data points (5 on each side of the line) centered on the \ion{Ne}{2} expected line location and used this as the peak line flux upper limit. We then used the average of the \uu{\ion{Fe}{2}} line widths in the upper limit calculation for \ion{Ne}{2}. The upper limit of the \ion{Ne}{2} line flux is shown in Table~\ref{tab:lines}.

We also checked to see if the \ion{Fe}{2} lines are spatially resolved like the \ion{Ar}{2} line from \citet{Worthen2024}. We did this by subtracting out a slice from the MRS data cube outside of the line from the slice at the peak of the line flux. This removes the resolved and unresolved emission from the dust and the central star, leaving only the emission from the \ion{Fe}{2} lines. The spatially resolved components of the \ion{Fe}{2} lines are shown in Fig.~\ref{fig:FeII_profiles}. We computed the spatially resolved line fluxes by extracting the spectra within the 3$\sigma$ red contours and outside of the JWST PSF FWHM shown in Fig.~\ref{fig:FeII_profiles}. We then fitted a Gaussian profile to the extracted line profile and integrated to get a resolved line flux. This was done on both sides of the disk and then summed together for each of the \ion{Fe}{2} lines. The spatially resolved line fluxes of the lines from the MRS data are shown in Table~\ref{tab:resolved_flx}.

\begin{figure*}
    \centering
    \includegraphics[scale=0.7]{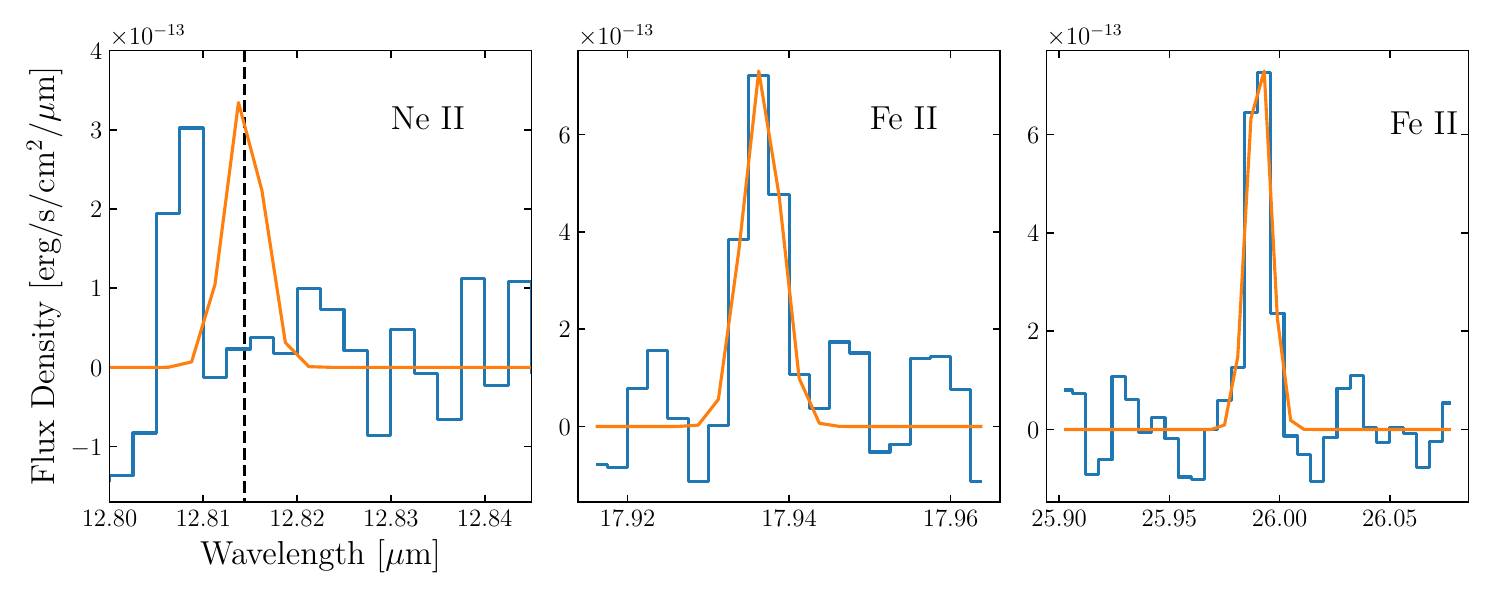}
    \caption{Left: MIRI MRS spectrum of the unresolved point source \bpic\ over a wavelength range that contains the \ion{Ne}{2} line at 12.81\,$\mu$m. The black dashed line shows the expected location of the line at the radial velocity of \bpic. The orange Gaussian is the corresponding line profile for the line flux upper limit we measure. Middle: \ion{Fe}{2} line at 17.94\,$\mu$m with best-fit Gaussian profile overlaid. Right: \ion{Fe}{2} line at 25.99\,$\mu$m with best-fit Gaussian profile overlaid. }
    \label{fig:Lines2}
\end{figure*}

\begin{figure*}
    \centering
    \includegraphics[scale=0.7]{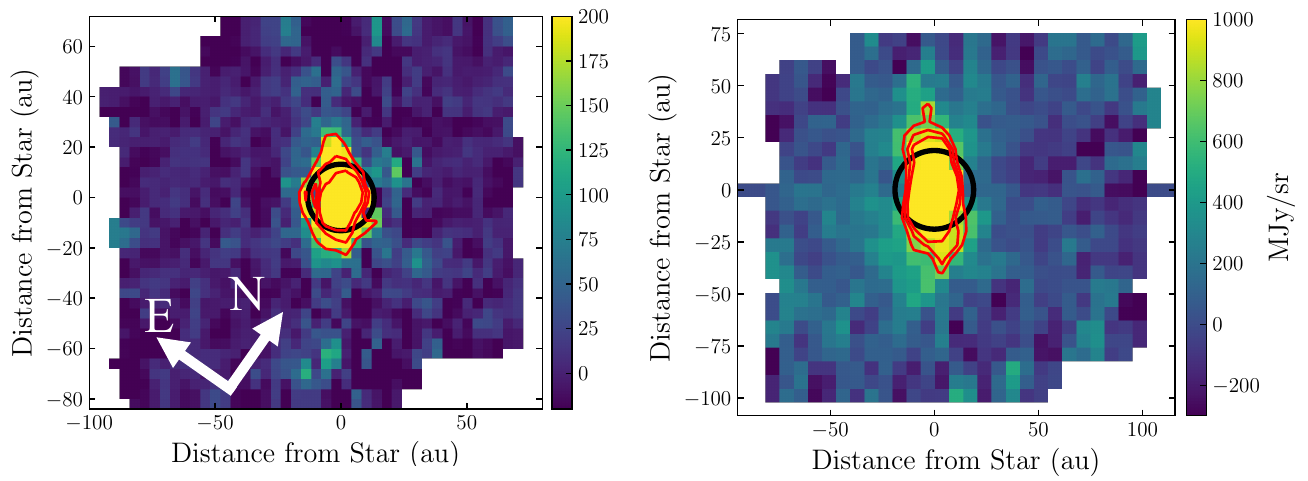}
    \caption{Left: PSF and continuum-subtracted MRS data cube slice at 17.94\,$\mu$m showing the spatially resolved structure of the \ion{Fe}{2} emission. The black circle shows the JWST PSF FWHM in the across slice direction at this wavelength and the red contours show 3, 5, and 10$\sigma$. Right: same as left but for the \ion{Fe}{2} line at 25.99\,$\mu$m. Note that the MRS Channel 4C PSF has a different FWHM in the across-slice and along-slice directions \citep[see][]{Law23}.}
    \label{fig:FeII_profiles}
\end{figure*}

\section{Nitrogen from HST/COS data}
\label{sec:N_data}

\begin{figure}
    \centering
    \includegraphics[width=0.48\textwidth]{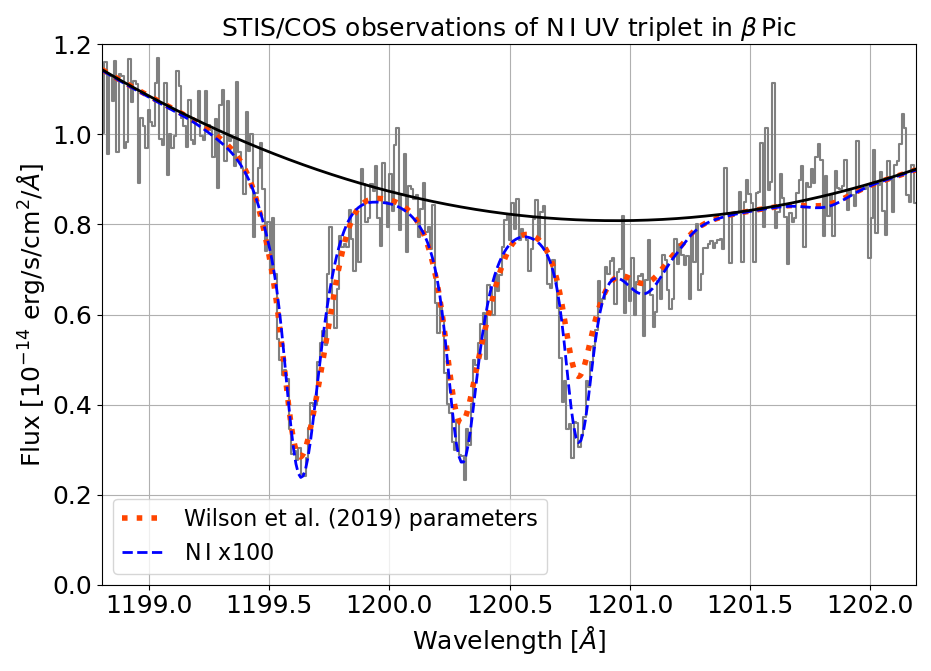}
    \caption{STIS/COS observations of the \ion{N}{1} triplet of UV resonance lines \citepalias{Wilson2019}. The data are consistent with
    the `stable' circumstellar N having a 100$\times$ overabundance relative to the refractory elements.}
    \label{fig:NI_triplet}
\end{figure}

Since constraining column densities from optically thick lines with complex profiles is challenging and often dependent on assumptions on the individual line components, we decided to have a closer look at the \citetalias{Wilson2019} constraints to see how a high column density of N$^{0}$ could reasonably be compatible with the data. \uu{P.~A.\ Str\o m (n\'e Wilson) kindly provided us with the reduced and co-added HST/COS data used in \citetalias{Wilson2019}.} When attempting to reproduce their results we discovered that their model spectrum (their Fig.~1) used an approximately 0.5$\times$ too narrow line-spread function (LSF). This can be seen in the convolved \ion{N}{1} lines of their Fig.~1, where their FWHM are closer to 10\kms\ rather than the expected 23\kms\ for the used instrument setting of COS. The consequence is that the column density of circumstellar \ion{N}{1} becomes underestimated. Using the G130M/1222 LSF from LP3 downloaded from the STScI website\footnote{\url{http://www.stsci.edu/hst/cos/performance/spectral\_resolution/}} (with FWHM $\approx$ 23\kms ), we find that many of the parameters used in \citetalias{Wilson2019} become more strongly correlated, e.g.\ the broadening parameter $b$ and column densities $N$ for the various absorption components. Using the broad LSF with the parameters found in \citetalias{Wilson2019} no longer produces a good fit (see Fig.~\ref{fig:NI_triplet}). Adjusting the ``stable circumstellar gas component'' CS$_0$ of \ion{N}{1} column density to a factor of 100$\times$ higher column density (i.e.\ $N_{\mathrm{N\,I}} = 10^{16.9}$\,cm$^{-2} = 7.9 \times 10^{16}$\,cm$^{-2}$) and fitting $b = 1.4$\kms\ (from previous $b = 1.0$\kms ), we get a much better match to the data (Fig.~\ref{fig:NI_triplet}). Our conclusion is that a N overabundance of 100$\times$ compared to refractory elements is indeed consistent with the data.

\end{document}